%% file: main.tex
\documentclass[conference]{IEEEtran}
\usepackage{cite}
\ifCLASSINFOpdf
  \usepackage[pdftex]{graphicx}
\else
\fi
\usepackage{multicol}
\usepackage{epstopdf}

\usepackage[cmex10]{amsmath}
\usepackage{amssymb,amsfonts}
\usepackage{amsthm}
\usepackage{booktabs}
\usepackage[tight,footnotesize]{subfigure}

\usepackage{fancyhdr} 
\begin{document}
\title{The Pointillist Family of Multitarget Tracking Filters}

\author{\IEEEauthorblockN{Roy Streit\IEEEauthorrefmark{2}, Christoph Degen\IEEEauthorrefmark{1}, Wolfgang Koch\IEEEauthorrefmark{1}}
\IEEEauthorblockA{\IEEEauthorrefmark{2} Metron, Inc., Reston, VA 02871, United States, 
	Email: streit@metsci.com}
\IEEEauthorblockA{\IEEEauthorrefmark{1} SDF Dept., Fraunhofer FKIE, Wachtberg, Germany, 
Email: \{Christoph.Degen, Wofgang.Koch\}@fkie.fraunhofer.de}}


%Telephone: (800) 555--1212, Fax: (888) 555--1212}


\maketitle
\thispagestyle{fancy}

\begin{abstract}
The family of pointillist multitarget tracking filters is defined to be the class of filters that is characterized by a joint target-measurement finite point process. The probability generating functional (PGFL) of the joint process is derived directly from the probabilistic structure of the tracking problem. PGFLs exemplify the analytic combinatoric method applied to the measurement to target assignment problems that are fundamental to the tracking problem. It is shown that multi-hypothesis tracking (MHT), joint probabilistic data association (JPDA), and many other now-classic tracking filters can be derived via PGFLs, and thus are members of the family of pointillist filters. When one or more of the target processes are dimensionally compatible, targets can be superposed. It is shown that the classic MHT filter for superposed targets is closely related to the multi-Bernoulli filter. A technique is presented for deriving the functional derivatives by ordinary differentiation, and both exact and approximate methods for evaluating the ordinary derivatives are discussed. 
\end{abstract}


\begin{IEEEkeywords}
	Multitarget tracking, Analytic combinatorics, Assignment problems, Finite point processes, Superposition	
\end{IEEEkeywords}

% \fi
\IEEEpeerreviewmaketitle

%\setcounter{tocdepth}{1}
%\tableofcontents

\setcounter{secnumdepth}{3}

\section{Introduction}
\input{introduction}

\section{Combinatorics and Tracking}
\label{sec:background}
\input{extendedIntroduction}

\section{Superposition and Marginalization of Finite Point Processes}
\label{sec:superposition}
\input{superpositionPlus}

\section{Notation and Models}
\label{sec:notation}
\input{background}

\section{Pointillist Filters without Superposition}
\label{sec:pointillistFiltersWithoutSuperposition}
\input{pointillistFiltersWithoutSuperposition}

\section{Pointillist Filters with Superposition}
\label{sec:pointillistFiltersSuperposition}
\input{pointillistFiltersWithSuperposition}

\section{Hybrid Pointillist Filters}
\label{sec:hybridPointillistFilters}
\input{hybridPointillistFilters}

\section{Closing the Bayesian Recursion}
\label{closingTheBayesianrecursion}
\input{closingBayesianRecursion}

\section{Target State Estimation}
\label{sec:commentsOnTargetSuperposition}
\input{targetStateEstimation}

\section{How to Design a Tracking Filter: An Engineer's Perspective}
\label{sec:howToDesignATrackingFilter}
\input{howToDesignATrackingFilter}

\section{Alternatives to Functional Differentiation}
\label{Alternatives to Functional Differentiation}
\input{alternativesToFunctionalDifferentiation}
\section{Conclusions And Future Work}
\label{sec:conclusionsAndFutureWork}
\input{conclusionsAndFutureWork}

\appendices
\section{Finite Point Process Basics}
\input{appendixA}

\section{Stacked Random Vectors and Random Finite Sets}
\input{appendixB}

\bibliographystyle{plain}
\bibliography{myBib}

\end{document}

%% file: introduction.tex
Classical methods of analytic combinatorics are used to characterize a large number of well known recursive, discrete-time, multitarget tracking filters. Filters that can be conceptualized in this way are called pointillist filters because they are built around finite point process models of the joint target-measurement process. Finite point processes are fully and uniquely characterized by probability generating functionals (PGFLs) \cite{Moyal} and, naturally, different tracking filters have different PGFLs. In tracking applications, the appropriate PGFL can often be derived directly from the modeling assumptions that underlie the tracking problem. PGFLs illuminate the similarities and differences between the combinatorial problems that underlie the filters. In other words, PGFLs are the basis of an ontology for the pointillist family of filters. This paper shows not only that many well known tracking filters are pointillist filters but also that the PGFL ontology can be used to derive new filters and bring to light previously unknown relationships between existing filters.\\
\indent Three classes of pointillist filters are identified. They are distinguished by their use of target superposition. One class does \textit{not} use target superposition. Individual target states are maintained by the filter, and different targets can have different state spaces. This class contains many well-known multitarget tracking filters, including Bayes-Markov \cite{Jazwinski}, MHT (multi-hypothesis tracking) \cite{ReidMht}, PDA (probabilistic data association) \cite{BarShalom2}, JPDA (joint PDA) \cite{BarShalom3}, PMHT (probabilistic multi-hypothesis tracking) \cite{StreitPMHT}, IPDA (integrated PDA) \cite{Musicki1}, and the JIPDA (joint IPDA) \cite{Musicki3} filters. In many instances the recognition that these filters can be formulated in terms of PGFLs is new to this paper. \\
\indent Another class \textit{does} use target superposition -- there is only one state space, shared by all targets, and separate target-specific states are not maintained by the filter. (Note that targets cannot be superposed if their state spaces are different.) Pointillist filters that use superposition include the PHD (probability hypothesis density) intensity \cite{Mahler03}, CPHD (cardinalized PHD) intensity \cite{Mahler07}, generalized PHD intensity \cite{ClarkGeneralizedPhd}, and the multi-Bernoulli intensity \cite{MahlerBook} \cite{VoMB1}  \cite{VoMB2} filters. Other filters in this class are superposed versions of the JPDA, the JIPDA, and the PMHT filters. The multi-Bernoulli filter is seen to be equivalent to the JIPDA filter with superposition, a fact that has also been studied in \cite{Williams}. \\
\indent A third class of pointillist filters is a hybrid class, in which targets are selectively superposed. Targets that are superposed have the same state space, but the other targets do not. Two hybridized versions of the PHD and generalized PHD filters are presented. The class of selectively hybridized filters is, to the knowledge of the authors, entirely new. \\
\indent The presentation of well-known, now classic, tracking filters in a unifying framework sheds new light on connections and differences between the filters. Moreover, it enables an engineer to design customized application-specific tracking filters using a palette of available point process target and measurement models to obtain possibly new and unique filters, matched to the requirement of the problem at hand. It is one of the purposes of this paper to show how an engineer can apply in practice the presented mathematical framework to the design of tracking filters.\\
\indent The paper also presents methods for computing functional derivatives \cite{Moyal} of the PGFL. These derivatives are used to obtain the probability distribution of the Bayes posterior point process and its summary statistics \cite{MahlerBook}, \cite{StreitPGFL}, \cite{ClarkGeneralizedPhd}. The method of secular functions \cite{StreitSecular} is important because it reduces functional differentiation to ordinary differentiation. The secular function is an analytic function of several complex variables, so its derivatives can be written as a multivariate Cauchy-style contour integral. The saddle point method (method of stationary phase) can be applied to approximate the contour integral \cite{Fusion2015Streit}. This strategy is fundamental to asymptotic analysis in analytic combinatorics \cite{Flajolet}. \\ 
\indent It is the completely undisputed and lasting contribution of Mahler to have introduced the notion of random finite sets into the target tracking community. Furthermore, he was the first who  derived -- among other contributions -- a popular multiple target tracker by using PGFLs within this framework. For several years, however, there has been an ongoing discussion in the tracking community of how to formulate the underlying mathematical structures and to link them to the classical mathematical theory of finite point processes that has been developed over decades.\\
\indent It is the express intent of this paper to bring different schools of the tracking community together by demonstrating that PGFLs are very precise and succinct models of the combinatorial probability structures involved in multitarget tracking, whether or not these models arise from a classical or a random set approach. The family of pointillist filters and the connections to analytic combinatorics were first proposed and discussed in \cite{ StreitFusion14Talk}. The first example of a tracking filter explicitly derived from a PGFL was the PHD intensity filter \cite{Mahler03}. The first application of PGFLs and the analytic combinatoric method to filters that do not employ target superposition was given for the PDA filter \cite{StreitPDA}.
\\ 
\indent The paper is organized as follows. Section \ref{sec:background} discusses the combinatorial assignment problems that are fundamental to multitarget tracking problems in the context of analytic combinatorics. Section \ref{sec:superposition} discusses superposition and marginalization of target processes in terms of PGFLs. General notation and models are presented in Section \ref{sec:notation}; to fix ideas, linear-Gaussian versions are provided in Section \ref{sec:targetMotionAndMeasurementModels}. Section \ref{sec:targetDetectionModeling} presents models for target detection, and Section \ref{sec:clutterModeling} discusses the clutter modelling.\\
\indent Section \ref{sec:pointillistFiltersWithoutSuperposition} (Section \ref{sec:pointillistFiltersSuperposition}) presents the class of pointillist filters that do not (do) employ target superposition. Hybrid pointillist filters are discussed in Section \ref{sec:hybridPointillistFilters}. Closing the Bayesian recursion and target state estimation are briefly discussed in Sections \ref{closingTheBayesianrecursion} and \ref{sec:commentsOnTargetSuperposition}, respectively. Section \ref{sec:howToDesignATrackingFilter} highlights the benefits of the paper for engineers. In particular a concrete non-standard example involving unresolved targets is presented to show how to model a tracking filter using PGFLs. Alternative methods for deriving functional derivatives---needed to derive the probability distribution and summary statistics of the Bayes posterior point process---are discussed in Section \ref{sec:alternativesToFunctionalDifferentiation}. Section \ref{sec:conclusionsAndFutureWork} gives conclusions and provides a table that overviews many of the filters discussed. Appendix \ref{sec:AppendixA} gives a brief overview of the basic concepts of finite point processes. Appendix \ref{sec:stackedVsRFS} shows that a stacked random vector differs from a random finite set defined as a realization of a finite point process.

%% file: extendedIntroduction.tex
Combinatoric problems are fundamental to multitarget tracking involving sensors whose output measurements are modeled as points. These points may belong to targets, or they may be spurious, that is, clutter measurements. Different applications impose different constraints on the choices, or assignments, of points to targets. Assignments that satisfy the constraints are said to be feasible. The prototype constraint is widely used ``at most one measurement per target rule" in which a target is assumed to generate at most one point measurement. Whatever the constraints are, the combinatoric problem is to model the feasible assignments in a probabilistic manner that supports target state estimation. \\
\indent The natural way to study combinatorial problems is to enumerate all feasible choices \cite{ReidMht}. This method is often impractical in large problems. Analytic combinatorics \cite{Flajolet} is a classical alternative method that has proved useful in many kinds of counting problems (e.g., how many ways can a convex polygon with $n+2$ sides be decomposed into $n$ non-overlapping triangles?). The basic strategy comprises two steps. The first and often  most difficult step is to discover the generating function (GF) that models the problem at hand. This step typically exploits the structure of the problem, e.g., a recursion, to find a closed form expression. The GF, once found, is an analytic function in the complex plane. The second step is to invoke analytic methods to study the GF to find expressions for the number of combinations. The behavior of GFs near their poles in the complex plane often determine asymptotic behavior \cite{Flajolet}. \\
\indent The PGFL approach is the analytic combinatoric method applied to the study of the multi-target tracking assignment problems. Specifically, GFs are replaced in tracking applications by generating functionals and, most often, PGFLs. \\
\indent Finding and exhibiting the PGFLs that model the various tracking filters in the pointillist family is the \textit{Discovery Step} of analytic combinatorics. Deriving the filter and closing the Bayesian recursion is the second \textit{Analytical Step}. These steps are discussed and used in Section \ref{sec:howToDesignATrackingFilter} to aid engineers in the design of tracking filters. In addition to Section \ref{sec:howToDesignATrackingFilter}, comments on closing the recursion are given in Section  \ref{closingTheBayesianrecursion} and when discussing specific filters. Implementation details and filter performance are not within the scope of the paper. Interested readers are referred to the literature.

%% file: superpositionPlus.tex
\indent It is assumed that readers have some familiarity with finite point processes and PGFLs in a general setting. The appendix of this paper gives a highly abbreviated presentation of the basic facts used in this paper. The authoritative texts on point processes are \cite{DaleyJonesVolume1} and \cite{DaleyJones}, and the fundamental paper on PGFLs is \cite{Moyal}. The first applications of PGFLs in a tracking context can be found in \cite{MahlerBook} and the papers \cite{Mahler03} and \cite{Mahler07}. More approachable expositions of relevant background material can be found in \cite{StreitPGFL} and \cite{StreitSecular}. References to specific filters are given where they are first discussed in technical detail. Better known filters are referenced to accessible textbooks wherein citations to the original literature can be found. \\
\indent Superposition and marginalization of finite point processes are fundamental concepts that are easily described in terms of the underlying PGFL. The case of several mutually independent processes is treated first since it is the most intuitive. \\
\indent Suppose that the PGFLs of $n\ge 1$ target processes are specified on the spaces $S_i, i=1,\dots,n$. These spaces need not be the same, so each target has its own test function. Denote the PGFL of process $i$ by $\Psi_i(h_i)$, where $h_i$ is a complex-valued test function defined on $S_i$, that is, $h_i:S_i\rightarrow \mathbb{C}$. Then the joint PGFL of the $n$ processes is defined by
\begin{equation}
\Psi(h_1,\dots,h_n)=\prod_{i=1}^{n}\Psi_i(h_i).\label{meanField}
\end{equation}
The product form of the joint PGFL $\Psi(h_1,\dots,h_n)$ holds if and only if the processes are mutually independent. When this is not the case, the PGFL must be specified in such a way as to account for the correlation between the processes. In any event, realizations of the joint process are Cartesian products of finite point sets in the spaces $S_i$. \\
\indent In the language of PGFLs, marginalizing over a point process is equivalent to setting its test function identically equal to one. Thus, the PGFL of the $i^{\textnormal{th}}$ marginal process is
\begin{equation}
\Psi_i(h_i)=\Psi(\dots,1,h_i,1,\dots).\label{marginalPGFL}
\end{equation}
Realizations of the $i^{\textnormal{th}}$ marginal process are finite point sets in the space $S_i$. When the processes are mutually independent, the PGFLs of the marginal processes are identical to the factors in the product (\ref{meanField}). \\
\indent The $n$ processes can be superposed if the spaces $S_i$ are copies of the same space, which is denoted by $S$ and referred to as the ground space. There is only one state space so there is only one test function. The PGFL of the superposed process is given by the "diagonal" of the joint PGFL, i.e., 
\begin{equation}
\Psi_S(h)\equiv\Psi(h,\dots,h),\label{superposedPGFL}
\end{equation}
where the test function of the superposed process $h(\cdot)$ is defined on $S$. Realizations of the superposed process are finite point sets in $S$. \\
\indent The filters of Section \ref{sec:pointillistFiltersWithoutSuperposition} do \textit{not} use superposition, and therefore have as many (possibly different) target state spaces and test functions as there are targets. The pointillist filters of Section \ref{sec:pointillistFiltersSuperposition} \textit{do} use superposition, and thus all targets share the same state space and there is only one test function. Finally, the hybrid pointillist filters discussed in \ref{sec:hybridPointillistFilters} superpose some targets and not others. The superposed targets share a state space but non-superposed targets do not. \\

%% file: background.tex
\indent In this section a consistent notation for the target and measurement models used in pointillist filters is presented. Specific probability distributions are not assumed. \\
\indent The initial reference time is denoted by $t_0$. The measurement sample times are denoted by $t_k$, $k=1,2,\dots\,$. It is assumed that $t_{k-1}<t_{k}$ for all $k$. The recursive time index is suppressed throughout for ease of presentation. \\
\indent Targets are assumed to move over time in a state space denoted by $S$, where $S\subset\mathbb{R}^{\dim(S)}$. Measurements are points in a measurement space $Y\subset \mathbb{R}^{\dim(Y)}$. \\
\indent The general models used for pointillist filters that do not superpose targets are discussed in this section. To introduce the meaning of the notation, equivalent versions for the standard linear-Gaussian Kalman filter \cite{KalmanFilter} are given in Section \ref{sec:KalmanFilter}. Point target and extended target detection models are discussed in Section \ref{sec:targetDetectionModeling}. Clutter modelling is described in Section \ref{sec:clutterModeling}.  

\subsection{Target Motion and Measurement Models}
\label{sec:targetMotionAndMeasurementModels}

\subsubsection{General Probability Models}
\label{sec:generalProbabilityModels}
\indent A target prior probability density function (PDF) is specified at the recursion start time $t_0\equiv t_{k-1}$. Six PDFs are needed:
\begin{itemize}
	\item $\mu_0(x_0)$, the (prior) target PDF at time $t_{k-1}$,
	\item $p_0(x|x_0)$, the Markovian target motion (transition) model from $x_0\in S$ at time $t_{k-1}$ to $x\in S$ at $t_k$, 
	\item $\mu(x)$, the predicted target PDF at time $t_k$, 
	\item $p(y|x)$, the likelihood function of a measurement $y\in Y$ at time $t_k$ conditioned on target state $x\in S$ at $t_k$,
	\item $p(y)$, the PDF of a measurement at time $t_k$ conditioned on the sequence of all measurements up to and including time $t_{k-1}$,
	\item $p(x|y)$, the Bayes posterior PDF at time $t_k$ conditioned on measurements up to time and including time $t_k$.
\end{itemize}
Three of these PDFs are determined by the others:
\begin{align}
	\label{predictedTargetPdf}
	\mu(x)=&\int_S \mu_0(x_0) p_0(x|x_0) \, dx_0 \\
	p(y)=&\int_S \mu(x) p(y|x) \, dx \\
	p(x|y)=&\,\frac{\mu(x)p(y|x)}{p(y)}.
\end{align}
The last expression is Bayes Theorem. \\

\subsubsection{Linear-Gaussian Models}
\label{sec:KalmanFilter}
\indent The general notation takes a familiar form for the classical linear-Gaussian Kalman filter \cite{KalmanFilter}. The target state space is $S=\mathbb{R}^{\dim(S)}$ and the measurement space is $Y=\mathbb{R}^{\dim(Y)}$. The time index is included in this subsection. Standard Kalman filter notation is adopted. Starting the recursion at time $t_{k-1}$, let
\begin{align}
\label{priorLG}
\mu_0(x_0)=\mathcal{N}(x_0;\hat{x}_{k-1|k-1},P_{k-1,k-1}),
\end{align}
where $\hat{x}_{k-1|k-1}\in S$ and $P_{k-1|k-1}\in\mathbb R^{\dim(S)\times\dim(S)}$ are the specified mean vector and covariance matrix, respectively, at time $t_{k-1}$. The target motion model is $x_k=F_{k-1} x_{k-1} + w_{k-1}$, where $F_{k-1}\in\mathbb R^{\dim(S)\times\dim(S)}$ is specified and the noise vector $w_k$ is Gaussian distributed with zero-mean and known covariance matrix $Q_{k-1}\in\mathbb R^{\dim(S)\times\dim(S)}$. The target motion model written in PDF form is
\begin{align}
p(x|x_0)=\mathcal{N}(x;F_{k-1} x_0,Q_{k-1}).
\end{align}
The predicted target PDF at time $t_k$ is 
\begin{align}
\mu(x)=\mathcal{N}(x;\hat{x}_{k|k-1},P_{k|k-1}),
\end{align} 
where the predicted mean and covariance matrix are given by 
\begin{align}
\hat{x}_{k|k-1}=&F_{k-1} \hat{x}_{k-1|k-1} \\
P_{k|k-1}=&F_{k-1} P_{k-1|k-1}F_{k-1}^T+Q_{k-1}.
\end{align}
The measurement model is $y_k=H_k x_k + v_k$, where the measurement matrix $H_k\in\mathbb R^{\dim(Y)\times\dim(S)}$ is specified and $v_k$ is an additive noise vector that is zero-mean and Gaussian distributed with known covariance matrix $R_k\in\mathbb R^{\dim(Y)\times\dim(Y)}$. The measurement likelihood function is 
\begin{align}
p(y|x)=\mathcal{N}(y;H_{k} x,R_{k}).
\end{align}
The predicted measurement PDF is 
\begin{align}
p(y)=\mathcal{N}(y;\hat{y}_{k|k-1},S_{k}),
\end{align}
where the predicted measurement and its covariance matrix are
\begin{align}
\hat{y}_{k|k-1}=\,&H_{k} \hat{x}_{k|k-1} \\
S_k=\,&H_k P_{k|k-1} H^{T}_k +R_k.
\end{align}
The measurement $y_k\in Y$ is given at time $t_k$. The innovation (residual measurement error) at time $t_k$ is defined by $\iota_{k}=y_k-\hat{y}_{k|k-1},$ so $S_k\in\mathbb R^{\dim(Y)\times\dim(Y)}$ is the innovation covariance matrix. The Kalman filter output is the posterior PDF  
\begin{align}
p(x)=\mathcal{N}(x;\hat x_{k|k},P_{k|k}),
\end{align}
where the updated mean and covariance matrix are
\begin{align}
\hat{x}_{k|k}=\,&\hat{x}_{k|k-1}+W_{k}\, \iota_{k}\\
P_{k|k}=\,&P_{k|k-1}-W_{k}S_{k}W^{T}_{k},\\
W_k =\,&P_{k|k-1}H^{T}_{k} S^{-1}_{k}.
\end{align}
The matrix $W_k\in\mathbb R^{\dim(S)\times\dim(Y)}$ is the filter gain matrix. \\

\subsection{Target Detection Modeling}
\label{sec:targetDetectionModeling}
\subsubsection{Targets With At Most One Point Measurement}
\indent Missed target detections are modeled by assuming that a target that is known to be present at state $x\in S$ at time $t_k$ is detected with probability $P^D_k(x)$, where $0\le P^D_k(x)\le 1$. The probability of missing the target detection is then $1-P^D_k(x)$. Suppressing the recursion time index, for all $x\in S$ let
\begin{align}
	\label{detectionprobs}
	a(x)\equiv1-P^D_k(x) \quad\text{and}\quad b(x)\equiv P^D_k(x).
\end{align}
The corresponding PGF is defined by
\begin{align}
	\label{GBMD}
	G_{M|x}^{\text{BMD}}(z) \equiv a(x) + b(x) z,
\end{align}
where $z\in\mathbb C$. Target detection probabilities may or may be not be the same for all targets, depending on the application.\\
\subsubsection{Extended Target Model}
An extended target is assumed to have a well defined state $x\in S$, e.g., an appropriately defined centroid. It is assumed that extended targets generate a random number $M\ge 0$ of independent and identically distributed (point) measurements in the space $Y$. The distribution of a target-originated measurement is taken to be the likelihood function $p(y|x)$. The number of measurements can also depend on the target state. The conditional PGF of the random variable $M$ is defined by
\begin{align}
	\label{extendedTargetPGF}
	G_{M|x}(z)\equiv\sum_{m=0}^\infty\,\Pr\text\{M=m\big{|}\,x\text\}\,z^m, 
\end{align}
where $\Pr\text\{M=m|\,x\text\}$ denotes the probability that $m$ measurements are generated by the extended target with state $x\in S$. The target is said to be detected if $M\ge 1$. For $M\ge1$, let 
\begin{align}
	\label{extendedTargetNumberOfMeasProbs}
	d_m(x)\equiv\Pr\{M=m\big{|}\,\text{target at state } x \text{ is detected}\},
\end{align}
so that $\sum_{m=1}^\infty d_m(x)=1$. Using the probabilities (\ref{detectionprobs}) gives 
\begin{align}
	G_{M|x}(z)=&\,a(x)+b(x)\sum_{m=1}^{\infty} d_m(x) z^m\\
	\equiv &\,a(x)+b(x)\,G_{D|x}(z),
\end{align}
where $G_{D|x}(z)$ is the PGF of the number of measurements generated by a detected target at $x$. It reduces to the ``at most one measurement per target'' model for $M\equiv 1$, that is, to $G_{M|x}^{\text{BMD}}(z)$ in (\ref{GBMD}) when $d_1(x)=1$ and $G_{D|x}(z)=z$. \\
%\indent When the context is clear, the PGFs $G_{M|x}(\zeta)$,  $G_{M|x}^1(\zeta)$  and $G_{D|x}(\zeta)$ are written more simply as $G_M(\zeta)$,  $G_{M}^1(\zeta)$ and $G_{D}(\zeta)$, respectively. 

\subsection{Clutter Modeling}
\label{sec:clutterModeling}
\indent The clutter process (also called the false alarm process) in the measurement space $Y$ is assumed to be either a non-homogeneous time-dependent Poisson point process (PPP) or a generalization called a cluster process. For all cluster processes, including PPPs, given the number of points, the points are i.i.d. (independently and identically distributed). The PGF (or $z$-transform as it is called in signal processing applications) of the probability distribution of the number of points plays a key role. PPPs are flexible, well understood, and widely used in diverse applications \cite{PPP}.\\ 
\indent PPP clutter is considered first. Denote the intensity function of the clutter process at time $t_k$ by $\lambda_k(y)$. The PPP is homogeneous if $\lambda_k(y)\equiv\text{constant}$ on $Y$; otherwise, it is non-homogeneous. Time dependence is suppressed in the notation, so that $\lambda_k(y)$ is written $\lambda(y)$. The mean number of clutter points in $Y$ is 
\begin{align}
	\label{meanNumberClutter}
	\Lambda\equiv\int_Y \lambda(y)\,dy.
\end{align}
To assure that $\Lambda$ is finite (and simplify the discussion), it is assumed that $Y$ is bounded. Define the clutter PDF $p_\Lambda(y)$ to be the normalized intensity function, 
\begin{align}
	\label{normalizedIntensity}
	p_\Lambda(y)\equiv\lambda(y)/\Lambda.
\end{align}
In this notation, the PGFL of the PPP clutter process is \cite{Karr}
\begin{align}
	\label{PGFLforClutter}
	\Psi_{C}^{\text{PPP}}(g)\equiv&\exp \bigg (-\Lambda+\Lambda \int_Y g(y) p_\Lambda(y) \, dy \bigg ),
\end{align} 
where the test function $g:Y \rightarrow \mathbb{C}$ is, like all test functions, bounded and integrable. \\
\indent The PGFL of clutter when modeled as a cluster process is     
\begin{align}
\label{pgflIIDClusterClutter}
\Psi_{C}^{\text{Cluster}} (g) \equiv G_C\left(\int_Y g(y)q(y)dy\right),
\end{align}
where $q(y)$ is the PDF of a clutter point $y\in Y$ and $G_C(z)$ is the PGF of the number of clutter points, namely, 
\begin{align}
	G_C(z) \equiv \sum_{c=0}^\infty \Pr\{C=c\} z^c,
\end{align}  
where $\Pr\{C=c\}$ denotes the probability that $c$ clutter measurements are generated. Note that (\ref{PGFLforClutter}) is a special case of (\ref{pgflIIDClusterClutter}) for $G_C(z)=\exp(-\Lambda+\Lambda z)$.

%% file: pointillistFiltersWithoutSuperposition.tex
\indent The joint PGFLs of several well-known recursive discrete-time tracking filters that do not superpose targets are exhibited in this section. The general notation of Section \ref{sec:notation} is used, and the recursive time index is suppressed. 

\subsection{Bayes-Markov Filter}
\label{sec:Bayes-MarkovFilter}
\subsubsection{Classical Problem}
The classical Bayes-Markov filter \cite{Jazwinski}, \cite{StoneStreitBMTT} is presented in the following. It assumes exactly one target to be always present. It is always detected, and it generates exactly one measurement. There is no clutter, i.e., spurious measurements unrelated to the target. Denote the target and measurement test functions by $h:S \rightarrow \mathbb{C}$ and $g:Y \rightarrow \mathbb{C}$, respectively. Both are bounded and integrable. The PGFL is
\begin{align}
	\label{BayesMarkov}
	\Psi_{\text{BM}}(h,g)\equiv\int_S \int_Y h(x) g(y) \mu(x) p(y|x) \,dy\,dx.
\end{align}
As a check, note that $\Psi_{\text{BM}}(1,1)\equiv\Psi_{\text{BM}}(h,g)|_{h(\cdot)=1,g(\cdot)=1}=1$, due to the fact that $\mu(x)$ and $p(y|x)$ are PDFs. 
\subsubsection{With Missed Detections}
Missed target detections are modeled using (\ref{detectionprobs}) \cite{Jazwinski}, \cite{StoneStreitBMTT}. %Define the PGFL of the target-originated measurement process conditioned on target in state $x$ by
%\begin{align}
%	\label{atMostOneMeasPerTargetPGFL}
%	\Psi_{T|x}(g)=a(x)+b(x) \int_Y  g(y) p(y|x) \,dy.
%\end{align}
The joint target-originated measurement process is 
\begin{align}
	\label{BayesMarkovMissedDet}
	\Psi&_{\text{BMD}}(h,g) \equiv \int_S h(x) \mu(x) G_{M|x}^{\text{BMD}}\left(\int_Y  g(y) p(y|x) \,dy\right)  dx\nonumber\\
	&=\int_S h(x) \mu(x) \bigg (a(x)+b(x) \int_Y  g(y) p(y|x) \,dy \bigg )  dx.
\end{align}
As a check, note that $\Psi_{\text{BMD}}(1,1)=1$. This expression shows up often in the sequel. It reduces to (\ref{BayesMarkov}) when the target detection probability is one.
\subsubsection{With Missed Detections and Extended Target}
The PGF (\ref{extendedTargetPGF}) of the number of measurements generated by the extended target with state $x\in S$ is assumed to be known. The PGFL for the extended target is \cite{Jazwinski}, \cite{StoneStreitBMTT}, \cite{MaskellPoissonModels}
\begin{align}
	\label{extendedBayesianTarget}
	\Psi&_{\text{BME}}(h,g)\equiv\int_S h(x) \mu(x) G_{M|x}\bigg ( \int_Y  g(y) p(y|x) \,dy \bigg ) dx.
\end{align}
If the target can generate at most one measurement, then $M\in\{0,1\}$ and $G_{M|x}(z)=a(x)+b(x)\,z$, $\ z\in \mathbb C$. Thus, (\ref{BayesMarkovMissedDet}) is a special case of (\ref{extendedBayesianTarget}).

\subsection{PDA Filter}
\subsubsection{Without Gating}
\label{pdaWithoutGating}
 In the PDA problem \cite{BarShalom2}, \cite{BarShalom3}, as in the classical Bayes-Markov problem, exactly one target is assumed to be present. However, the target may or may not be detected, and clutter may be also present. Realizations of the clutter process and the target measurement process are superposed in the sensor measurement space $Y$. The target measurement and clutter processes are assumed to be independent. The PGFL of superposed independent processes is the product of the PGFLs of the constituent processes \cite{HanbookOfStatistics19}. The target-originated measurement PGFL is given by (\ref{BayesMarkovMissedDet}) and the clutter PGFL is  (\ref{PGFLforClutter}), so the product    
\begin{align}
	\label{PDAnoGating}
	\Psi&_{\text{PDA}}^{\text{noGate}}(h,g)=\Psi_{\text{BMD}}(h,g) \, \Psi_{C}^{\text{PPP}}(g)\nonumber\\
	&\quad = \int_S h(x) \mu(x) \bigg (a(x)+b(x) \int_Y  g(y) p(y|x) \,dy \bigg )  dx\nonumber\\
	&\quad \quad\times \exp \bigg (-\Lambda+\Lambda \int_Y g(y) p_\Lambda(y) \, dy \bigg ).
\end{align}
is the PGFL of the superposed processes.

\subsubsection{With Gating} 
\label{sec:PDAWithGate}
The PGFL is modified to accommodate gating. The gate, denoted by $\Gamma$, is a specified subset of $Y$. The probability that a target-originated measurement falls within the gate is 
\begin{align}
	\label{gateProbability}
	P_\Gamma\equiv\Pr\{y\in\Gamma\}=\int_\Gamma p(y)\,dy.
\end{align}
The gate can be chosen arbitrarily at time $t_k$ so long as $\Pr\{y\in\Gamma\}\ne 0$; however, it is typically chosen so $\Pr\{y\in\Gamma\}$ is near one. Gating the PPP clutter process to $\Gamma$ yields a PPP \cite{PPP} whose PGFL is given by
\begin{align}
	\label{PGFLforGatedClutter}
	\Psi_{C}^{\text{PPPgated}}(g)=&\exp \bigg (-|\Gamma|+|\Gamma| \int_\Gamma g(y) p_{\Gamma}(y) \, dy \bigg ),
\end{align}
where the expected number of clutter points in the gate is 
\begin{align}
	\label{expectedNumberOfClutterPoints}
	|\Gamma|\equiv \int_\Gamma \lambda(y)\,dy
\end{align}
and the clutter PDF is the intensity normalized by gate volume:
\begin{align} 
	\label{clutterPDFGated}
	p_\Gamma(y)=
	\left\{
		\begin{array}{ll}
			\lambda(y)/|\Gamma|, & \mbox{if } y\in \Gamma \\
			0, & \mbox{if } y \notin \Gamma.
		\end{array}
	\right.
\end{align}
Censoring target measurements lying outside $\Gamma$ gives (cf. (\ref{BayesMarkovMissedDet}))
\begin{align}
	\label{PDAwithMissedDet}
	\Psi&_{\text{BMD}}^\text{gated}(h,g) \nonumber\\
	&=\int_S h(x) \mu(x) \bigg (a_\Gamma(x)+b_\Gamma(x) \int_\Gamma  g(y) p_\Gamma(y|x) \,dy \bigg )  dx, \nonumber
\end{align}
where the probability that a target at $x\in S$ is detected \textit{within} the gate is 
\begin{align}
	b_\Gamma(x)\equiv b(x)\,P_\Gamma,
\end{align}
so that $a_\Gamma(x)\equiv 1-b(x)P_\Gamma$ is the probability that it is not detected in the gate, and the conditional PDF of a gated target-originated measurement is
\begin{align}
	\label{gatedMeasurementPDF}
	p_\Gamma(y|x)=
		\left\{
		\begin{array}{cl}
		(P_\Gamma)^{-1}\,p(y|x), & \mbox{if } y\in \Gamma \\
		0, & \mbox{if } y \notin \Gamma.
		\end{array}
	\right.
\end{align}
Using these expressions gives 
\begin{align}
	\label{PDAwithGating}
	\Psi&_{\text{PDA}}(h,g)\equiv\Psi_{\text{BMD}}^{\text{gated}}(h,g) \, \Psi_{C}^{\text{PPPgated}}(g)\nonumber\\
	&=\int_S h(x) \mu(x) \bigg (1-b(x)P_\Gamma+b(x) \int_\Gamma  g(y) p(y|x) \,dy \bigg )  dx\nonumber\\
	&\quad\times\exp \bigg (-|\Gamma|+|\Lambda| \int_\Gamma g(y) p_\Lambda(y) \, dy \bigg ).
\end{align}
This expression holds only for $y\in\Gamma$. The gated PGFL reduces to the PGFL for ungated measurements when the gate is the entire measurement space, i.e., $\Gamma=Y$.

\subsubsection{Extended Target} Given the PGF (\ref{extendedTargetPGF}) of the number of measurements generated by the extended target, the PGFL for the ungated extended target is the product 
\begin{align}
\label{PDAwithExtendedTarget}
\Psi_{\text{PDAE}}&(h,g)=\Psi_{\text{BME}}(h,g)\,\Psi_{C}^{\text{PPP}}(g)\nonumber\\
=&\int_S h(x) \mu(x) G_{M|x}\bigg ( \int_Y  g(y) p(y|x) \,dy \bigg ) dx \nonumber\\
&\times \exp \bigg (-\Lambda+\Lambda \int_Y g(y) p_\Lambda(y) \, dy \bigg ).
\end{align}
Gating the extended target measurements requires counting the number of ways that $m$ target-originated measurements can fall within the gate. It is not considered further here.  

\subsection{JPDA Filter}
Exactly $n\ge 1$ targets are assumed to be present. Clutter is assumed to be present. Target measurement and clutter processes are assumed to be mutually independent. Gating can be accommodated in the manner outlined below for the JIPDA filter in Section \ref{sec:JIPDAGating}, so it is not done here.\\
\indent The targets may or may not be detected. If detected, a target generates at most one measurement. All measurements generated are superposed in the same measurement space $Y$. Which measurements are target-originated and which are clutter is unknown. The PGFL of the $i^\textnormal{th}$ target measurement process is given by
\begin{align}
	\label{BayesMarkovMissedDetTargetNumberi}
	\Psi&_{\text{BMD}(i)}(h_i,g)\nonumber\\
	&=\int_{S_i} h_i(x) \mu_i(x) G_{M_i|x}^{\text{BMD}}\left(\int_Y  g(y) p(y|x) \,dy\right)\,dx\\
	&=\int_{S_i} h_i(x) \mu_i(x) \bigg (a_i(x)+b_i(x) \int_Y  g(y) p_i(y|x) \,dy \bigg )  dx,\nonumber
\end{align}
where the quantities in this expression are the same as in (\ref{BayesMarkovMissedDet}) but with an index $i$ added to make them specific to the $i$th target. The state spaces $S_i$ can be different or they can be copies of the same space. \\
\indent Target processes are not superposed in a common state space. Consequently, a different test function $h_i(x):S_i\rightarrow \mathbb{R}$ is necessary for each target. In contrast, the same measurement test function $g$ is used because all measurements, whether target-originated or clutter, are superposed in the measurement space $Y$. \\
\indent Because clutter and target measurement processes are mutually independent, the joint PGFL is the product of PGFLs:
\begin{align}
	\label{PGFLforJPDA}
	\Psi_{\text{JPDA}}(h_1,\dots,h_n,g)=\Psi_{C}^{\text{PPP}}(g)\prod\limits_{i=1}^{n}\Psi&_{\text{BMD}(i)}(h_i,g),
\end{align}  
where $\Psi_{C}^{\text{PPP}}(g)$ is given by (\ref{PGFLforClutter}). \\
\indent The PGFL of the marginal process for the $j$th target, $j=1,\dots,n$, is found by setting $h_i(x)=1$ for $i\ne j$. Explicitly,
\begin{align}
	\Psi_{\text{JPDA}(j)}(h_j,g)=&\,\Psi_{C}^{\text{PPP}}(g)\Psi_{\text{BMD}(j)}(h_j,g)\cdot\nonumber \\
	&\prod\limits_{i=1,\,i\ne j}^{n}\Psi_{\text{BMD}(i)}(1,g).
\end{align}
The marginal PGFLs are used for approximation purposes. They do not characterize the full JPDA distribution. For the now-standard derivation of the JPDA filter see \cite{BarShalom2}, \cite{BarShalom3}. 

\subsection{PMHT Filter}
\label{sec:PMHT}
The PMHT filter and enhanced versions were published in \cite{StreitPMHT}, \cite{StreitPMHT1}, \cite{StreitPMHT2}. It is assumed that there are exactly $n\ge 1$ targets present. Target-originated measurements are modeled as PPPs that are conditioned on target state $x\in S$. The mean number of measurements generated by the $i^{\textnormal{th}}$ target is denoted by $\Lambda_i(x)$, and the conditional likelihood is $p_{i}(y|x)$. The PGF of the $i^{\textnormal{th}}$ target-originated measurement process conditioned on a target at $x \in S$ is (\textit{cf}. (\ref{extendedTargetPGF}))
%\begin{align}
%\label{PGFLforTargeti}
%\Psi_{T_i|x}^{\text{PMHT}}(g)=&\exp \bigg (-\Lambda_i(x)+\Lambda_i(x) \int_Y g(y) p_{i}(y|x) \, dy \bigg ),
%\end{align} 
\begin{align}
	G_{M_i|x}^{\text{PPP}}(z)\equiv \exp \big(-\Lambda_i(x)+\Lambda_i(x) z \big).\label{PMHTmeasPPP}
\end{align}
Although not formulated in terms of PGFLs, Poisson models for the number of measurements were first incorporated into the PMHT and histogram PMHT filters in \cite{Davey}, \cite{Davey2}.\\
\indent The PGFL of the $i^{\textnormal{th}}$ target is 
\begin{align}
&\Psi_{\text{BMD}(i)}^{\text{PPP}}(h_i,g)\nonumber\\
&=\int_S h_i(x) \mu_i(x) G_{M_i|x}^{\text{PPP}}\left(\int_Y g(y) p_{i}(y|x) \, dy\right)\,dx \nonumber \\
&=\int_S h_i(x) \mu_i(x) \exp\bigg (-\Lambda_i(x)+\Lambda_i(x)\int_Y g(y) p_i(y|x)\,dy \bigg ).
\end{align}
This expression differs from the JPDA model (\ref{BayesMarkovMissedDetTargetNumberi}) only in the choice of the conditional PGFL of the measurements. The $i^{\textnormal{th}}$ target model reduces to the extended target measurement model (\ref{extendedBayesianTarget}) when the number of measurements is Poisson distributed with mean $\Lambda_i(x)$, that is, when $G_{M|x}(z)\equiv\exp \left(-\Lambda_i(x)+\Lambda_i(x) z \right)$. \\
\indent Targets are mutually independent by assumption, so the joint PGFL of the PMHT filter is the product of the target-specific PGFLs,
\begin{align}
	\label{PGFLforPMHT}
	\Psi_{\text{PMHT}}(h_1,\dots,h_n,g)=\prod\limits_{i=1}^{n}\Psi_{\text{BMD}(i)}^{\text{PPP}}(h_i,g).
\end{align} 
As with JPDA, different targets have different test functions. PMHT, unlike JPDA, models clutter using a ``diffuse" target, that is, as a target with a high measurement variance. \\
\indent The PGFL (\ref{PGFLforPMHT}) is, by inspection, a linear functional (see the discussion following (\ref{linearFnal})) in the test functions $h_1,\dots,h_n$. From the discussion in Appendix \ref{LinearMomentsandPDFS}, it follows that realizations of the joint target point process have, with probability one, exactly one target in each state space, and that the Bayes posterior PDF and the intensity functions are identical. \\
\indent Denote a realization by $x_i\in S_i$, $i=1,\dots,n$. It can be shown that the joint posterior PDF is proportional to the product (over the measurements) of a probabilistic mixture of measurement likelihoods. The $n$ mixing proportions are 
\begin{align}
	\pi_i(x_i) =\Lambda_i(x_i)\bigg / \sum_{j=1}^n \Lambda_j(x_j),\quad\text{ for }i=1,\dots,n.
\end{align}
This result is reasonable given the well-known relationship between Poisson mixtures and the multinomial distribution \cite{Kingman}, \cite{JohnsonKotz}, \cite{Grim}. Details are straightforward and are omitted. The original PMHT filter assumes that the spatial rates are constant, $\Lambda_i(x)\equiv\Lambda_i$, and applies the EM method to find point estimates for the $n$ targets and other parameters. \\

\subsection{IPDA Filter}
\label{sec:IPDA}
The IPDA presented in \cite{Musicki1}, \cite{Musicki2}, \cite{MusickiEvans1994} \cite{ChallaMorelandeMusickiEvans}, \cite{MusickiSongStreit} shares all the assumptions of the PDA except for one --- it assumes that \textit{at most} one target exists. Let $\chi_0$, $\,0\le\chi_0\le 1$, denote the initial probability that the target exists, i.e., that there is $N=1$ target. Let $\chi$ denote the updated probability of target existence at current time (see \cite{Musicki1} for details). The PGF of the number of targets at the current time is therefore
\begin{align}
G_N(z)=1-\chi+\chi \,z.\label{PGFforTargetNumber}
\end{align}  
The PGFL for the target-generated measurement process is 
\begin{align}
	\Psi_{\text{IPDA}}(h,g)&=G_N(\Psi_{\text{BMD}}(h,g))\nonumber \\
	&=1-\chi+\chi\,\Psi_{\text{BMD}}(h,g),
\end{align} 
where $\Psi_{\text{BMD}}(h,g)$ is given by (\ref{BayesMarkovMissedDet}). Superposing an independent clutter process gives the PGFL for IPDA as  
\begin{align}
	\label{IPDA}
	\Psi_{\text{IPDA}}(h,g)=\big ( 1-\chi+\chi\,\Psi_{\text{BMD}}(h,g) \big )\,\Psi_{C}^{\text{PPP}}(g),
\end{align}
where the clutter PGFL $\Psi_{C}^{\text{PPP}}(g)$ is given by (\ref{PGFLforClutter}). \\
\indent In \cite{ChallaVoWang2002} the IPDA filter is derived using random sets. In \cite{MusickiLaScala2008} and \cite{MusickiEvans2002} further versions of the IPDA filter are presented. These investigations may represent another class of pointillist filters, but they are outside the scope of this paper.

\subsection{JIPDA Filter}
\subsubsection{Without Gating}
\label{sec:JIPDA}
The JIPDA filter was presented first in \cite{Musicki3} (see also  \cite{ChallaMorelandeMusickiEvans}, \cite{MusickiSongStreit}). The JIPDA process assumes that at most $n\ge 1$ targets exist. Let $\chi_i$ denote the existence probability for target $i$, i.e., the probability that target $i$ is present. The target-originated measurement process for target $i$ is, using the same notation as in (\ref{BayesMarkovMissedDetTargetNumberi}),
\begin{align}
	\label{pgflJIPDAOneTarget}
	\Psi_{\text{IPDA}(i)}(h_i,g)=1-\chi_i+\chi_i\,\Psi_{\text{BMD}(i)}(h_i,g).
\end{align}
Different test functions $h_i$ are needed because the each target has its own state space. Assuming the target-originated measurement processes are mutually independent of each other and of the clutter process gives the PGFL for JIPDA as 
\begin{align}
	\label{pgflForJIPDAprocess}
	\Psi_{\text{JIPDA}}&(h_1,\dots,h_n,g)\nonumber
	=\Psi_{C}^{\text{PPP}}(g)\prod_{i=1}^n \Psi_{\text{IPDA}(i)}(h_i,g) \nonumber\\
	=&\,\Psi_{C}^{\text{PPP}}(g)\prod_{i=1}^n \bigg (1-\chi_i+\chi_i\,\Psi_{\text{BMD}(i)}(h_i,g)\bigg ).
\end{align}
As checks, note that $\Psi_{\text{JIPDA}}(1,\dots,1)=1$ and that $\Psi_{\text{JIPDA}}(\cdot)$ reduces to $\Psi_{\text{JPDA}}(\cdot)$ for $\chi_i=1$.\\
\indent Analogously to the JPDA, the PGFL of the marginal process for the $j$th target, $j =1,...,n$ can be determined by setting $h_i(x) = 1$ for $i\neq j$. It is given by
\begin{align}
		\label{marginalJIPDA}
		\Psi&_{\text{JIPDA}(j)}(h_j,g) \nonumber\\
		&\,\, =\Psi_{C}^{\text{PPP}}(g) \bigg (1-\chi_j+\chi_j\,\Psi_{\text{BMD}(j)}(h_j,g)\bigg )\nonumber\\ 
		&\,\,\,\,\,\,\,\times \prod_{i=1,i\neq j}^n \bigg (1-\chi_i+\chi_i\,\Psi_{\text{BMD}(i)}(1,g)\bigg ).
\end{align}
The marginal defined in (\ref{marginalJIPDA}) can be used in various ways to approximate the joint distribution and thereby close the Bayesian recursion.\\
\indent In \cite{MorelandeChalla}, a closely related version of the JIPDA is presented using random sets. The versions differ only in the method used to start new tracks.

\subsubsection{With Gating}
\label{sec:JIPDAGating}
Gating is incorporated in a manner analogous to that of Section \ref{sec:PDAWithGate} for the PDA. Denote the measurement gate of the $i^\textnormal{th}$ target by $\Gamma_i\subset Y$, $i=1,...,n$. Let $\Gamma \equiv \cup_{i=1}^n\Gamma_i$. Assuming PPP clutter, the expected number of clutter points, $|\Gamma|$, in $\Gamma$ is given by (\ref{expectedNumberOfClutterPoints}). The corresponding clutter PDF $p_\Gamma(y)$ is given by (\ref{clutterPDFGated}). The gated clutter process is a PPP, and its PGFL is 
\begin{align}
	&\Psi_{C}^{\text{PPPgated}}(g) =\exp\Big(-|\Gamma| + |\Gamma|\int_{\Gamma}g(y)p_{\Gamma}(y)dy\Big).
\end{align}
Let $P_{\Gamma_i}$ denote the probability that a target-originated measurement falls within the gate of the $i$th target, that is 
\begin{align}
\label{gateProbabilityNTargets}
P_{\Gamma_i}\equiv\Pr\{y\in\Gamma_i\}=\int_{\Gamma_i} p(y)\,dy.
\end{align}
Then, the PGFL of the joint target-measurement process for the $i$th target is given by
\begin{align}
	\Psi&_{\text{BMD}(i)}^{\text{gated}}(h_i,g) \nonumber\\
	&=\int_S h(x) \mu(x) \bigg (a_{\Gamma_i}(x)+b_{\Gamma_i}(x) \int_{\Gamma_i}  g(y) p_{\Gamma_i}(y|x) \,dy \bigg ),
\end{align}
where the probability that the $i$th target at $x\in S_i$ is detected \textit{within} the gate is 
\begin{align}
b_{\Gamma_i}(x)=b(x)\,P_{\Gamma_i},
\end{align}
so that $a_{\Gamma_i}(x)=1-b(x)P_{\Gamma_i}$ is the probability that it is not detected in the gate, and the conditional PDF of a gated target-originated measurement is
\begin{align}
\label{gatedMeasurementPDFJIPDA}
p_{\Gamma_i}(y|x)=
\begin{cases}
	(P_{\Gamma_i})^{-1}\,p(y|x), & \mbox{if } y\in \Gamma_i \\
	0, & \mbox{if } y \notin \Gamma_i,
\end{cases}
\end{align}
which means that each target has its own gate and its likelihood function is normalized, so that it is a PDF within this gate. Therefore, the PGFL of the Bayes posterior process of the gated JIPDA filter is given by
\begin{align}
	\label{pgflForJIPDAprocessGated}
	&\Psi_{\text{JIPDA}}^{\text{gated}}(h_1,\dots,h_n,g)\nonumber\\
	&\,\,\,=\,\Psi_{C}^{\text{PPPgated}}(g)\prod_{i=1}^n \bigg (1-\chi_i+\chi_i\,\Psi_{\text{BMD}(i)}^{\text{gated}}(h_i,g)\bigg ).
\end{align}
The PGFL of the marginal process for the JIPDA including gating is determined analogously to Section \ref{sec:JIPDA}.

\subsection{MHT Filter}
\label{JIPDAandMHT}
In the standard Multi-Hypothesis Tracking (MHT) algorithm, presented in \cite{ReidMht}, the filter \textit{generates a set of data association hypotheses to account for all possible origins of every measurement.} In a specific iteration the hypothesis are generated, that a certain target is a false alarm, belongs to an existing target, or occurs due to a new target. Based on the target states various techniques for reducing the numerical complexity like gating, merging, pruning were proposed by the community. However, there is only one un-pruned and complete set of MHT hypothesis. \\
\indent Assume that the JIPDA filter derived from (\ref{pgflForJIPDAprocess}) is extended by a process that allows data induced targets. Then, assuming $m$ measurements were received it is given by 
\begin{align}
\label{pgflForJIPDADataDriven}
& \Psi_{\text{MHT}}(h_1,\dots,h_{n+m},g)  \nonumber\\
&\,\,\,=  \Psi_{C}^{\text{PPP}}(g)\prod_{i=1}^n \bigg (1-\chi_i+\chi_i\,\Psi_{\text{BMD}(i)}(h_i,g)\bigg)\nonumber\\
&\,\,\,\,\,\,\,\times \prod_{j=1}^m \bigg (1-\gamma_j+\gamma_j\,\Psi_{\text{BMD}(j)}^{\text{Data}}(h_{n+j},g)\bigg),
\end{align} 
where $\gamma_j$ denotes the probability that measurement $j$ was generated by a target that is not modeled yet, and 
\begin{align}
&\Psi_{\text{BMD}(j)}^{\text{Data}}(h_j,g)\nonumber\\
&\equiv\int_S h_j(x) \xi_j(x) G_{M|x}^{\text{BMD}}\left(\int_Y  g(y) p_j(y|x) \,dy\right)  dx
\end{align}
is a data-driven PGFL that assumes that this new target has (specified) \textit{a priori} PDF $\xi_j(x)$ and generated a measurement with conditional PDF $p_j(y|x)$. The test functions $h_{n+1},...,h_{n+j}$ correspond to the $m$ data-induced PGFLs. \\ 
\indent The PGFL given by the data-driven JIPDA in (\ref{pgflForJIPDADataDriven}) captures the complete un-pruned set of all hypothesis of a single scan MHT algorithm. The correct Bayesian way to derive the target state estimates is to apply marginalization over all except one target state, exactly as it is done in the JIPDA filter. \\ 
\indent To keep the numerical complexity feasible, many approximation methods (gating, merging) and decision techniques (pruning) can be applied to close the Bayesian recursion. These considerations are outside the scope of this paper.\\
\indent The close relationship between the IPDA and the MHT filters was first noted in \cite{MusickiEvans2005}. However, the connection was not discussed there in terms of PGFLs. The connection between MHT, JIPDA, and multi-Bernoulli filters has been studied using the framework of random finite sets in \cite{Williams}.\\
\indent The study of a multiple scan MHT filter, i.e., an MHT filter that keeps measurement origin hypotheses for several successive scans, has already been studied and will be presented in a further publication.\\

%% file: pointillistFiltersWithSuperposition.tex
\indent In this section the PGFLs of filters with target superposition are presented, that is all targets share the same target state space.\\
\indent The general notation of Section \ref{sec:notation} is used throughout, and the recursive time index is suppressed. 
\subsection{Superposition in JPDA and Other Multitarget Filters}
\label{sec:JPDAS}
The multitarget processes in Section \ref{sec:pointillistFiltersWithoutSuperposition} are such that each target had its own state space. When the state spaces are copies of the same space, then it is possible to superpose the target processes into one process on the common space. It is seen in this section that it is straightforward to superpose targets once the joint multitarget PGFL is known. In the following it is shown how three originally non-superposed filters can be superposed.\\
\indent One problem with superposition is the loss of target-specific labels. As noted in Section \ref{sec:commentsOnTargetSuperposition}, estimating target-specific PDFs from the superposed process requires separate procedures that are replete with their own special difficulties. \\
\indent Another problem, noted below in Section \ref{closingTheBayesianrecursion}, is that it mis-models multiple targets. The PDF $\mu(\cdot)$ plays a more significant role for superposed targets than it does for one target. This is because the target states are (typically) i.i.d. samples from a random variable whose PDF is $\mu(\cdot)$. The mis-modeling would disappear only if the samples could somehow be drawn without replacement from different targets. 
   
\subsubsection{JPDA}
The joint probability distribution in JPDA is marginalized over all but one of the $n$ targets to obtain a target-specific PDF. When the target state spaces are all the same, i.e., $S_i\equiv S$ for all $i$, target point processes can be superposed instead of marginalized. The PGFL of the JPDA with superposition (JPDAS) filter is found by substituting 
\begin{align}
	\label{targetSuperposition}
	h_i(\cdot)\equiv h(\cdot)
\end{align}
into the PGFL (\ref{PGFLforJPDA}) for the JPDA. The resulting PGFL is    
\begin{align}
	\label{PGFLforJPDAS}
	\Psi_{\text{JPDAS}}&(h,g)=\Psi_{\text{JPDA}}(h,\dots,h,g)\nonumber\\
	&=\Psi_{C}^{\text{PPP}}(g)\prod\limits_{i=1}^{n}\Psi_{\text{BMD}(i)}(h,g).
\end{align}
In JPDAS, as in JPDA, targets can have different motion models, measurement models, and detection probabilities. Note too that the superposition procedure can be limited so that it includes some targets and not others. 
\subsubsection{JIPDA, PMHT, and Other Filters}
\indent Targets can be superposed in other joint multitarget point processes with known PGFLs when identical target state spaces are assumed. The PGFL of the superposed process is found by substituting (\ref{targetSuperposition}). For example, from (\ref{pgflForJIPDAprocess}) and (\ref{PGFLforPMHT}), 
\begin{align}
	\label{SuperposedJIPDAS}
	\Psi_{\text{JIPDAS}}(h,g)=&\Psi_{\text{JIPDA}}(h,\dots,h,g)\\
	\label{SuperposedPMHT}
	\Psi_{\text{PMHTS}}(h,g)=&\Psi_{\text{PMHT}}(h,\dots,h,g)
\end{align}
are, respectively, the PGFLs for JIPDA and PMHT with superposition. 
 
\subsection{PHD Intensity Filter} 
\label{sec:phdIntensityFilter}
The probability hypothesis density (PHD) filter is proposed in \cite{MahlerBook} and the papers \cite{Mahler03} and \cite{Mahler07}. The intensity filter (iFilter) is proposed in \cite{PPP}. The difference between these filters lies in their state space. While the PHD filter uses the state space $S$ (typically $S=\mathbb R^d$, $d>0$), the iFilter uses an augmented state space $S^+ = S\cup\phi$, where $S$ is equal to the state space of the PHD filter and $\phi$ is the hypothesis space, used to model clutter by scatterers. By a careful choice of certain parameters the Bayes posterior of the iFilter on $S^+$ is the same as the Bayes posterior of the PHD filter. For details see \cite{StoneStreitBMTT}. A detailed comparison of both filters using the PGFL derivation can be found in \cite{StreitPGFL}.\\
\indent Targets are assumed to have the same state space $S$ and are superposed. It is assumed that the targets constitute a finite point process at the recursion start time. This is the prior process at the initial (start up) time, but thereafter it is the Bayes posterior point process. The PGFL of the process, denoted by $\Psi_0(h,g)$, is typically not a PPP, so it is approximated by a PPP to close the Bayes recursion. To this end, the joint conditional PDF is approximated by the product of the marginal PDFs. Details can be found in \cite{MahlerBook} and \cite{PPP}.\\
\indent It is important that the predicted target process be a PPP. This is guaranteed to be the case assuming independent probabilistic thinning (target death), and assuming independent Markovian targets having the same motion model $p_0(x|x_0)$ of Section \ref{sec:generalProbabilityModels}. It also holds if a new-target (birth) process is superposed, provided the birth process is a PPP independent of the target process. \\
\indent The details of the predicted target PPP are of little concern here -- it suffices to assume that it is a (recursively) specified PPP with intensity $\overline N \,\mu(x)$, where the predicted number of targets is $\overline N\equiv E[N]$, $N$ being the random variable modeling the number of targets, and $\mu(x)$ is the predicted target PDF. The mean $\overline N$ and PDF $\mu(x)$ are determined by details of the prediction process. (In particular, note that $\mu(x)$ is not given by (\ref{predictedTargetPdf}) except in special cases.)\\
\indent Since $N$ is Poisson distributed with mean $\overline N$, its PGF is (when there is at most one target, see (\ref{PGFforTargetNumber})) 
\begin{align}
	\label{pgfPoissonTargets}
	G_{N}^{\text{PPP}}(z)=e^{-\overline N+\overline N z}.
\end{align}  
The predicted target states are i.i.d. by PPP assumption and are drawn from the PDF $\mu(x)$. The target measurement functions and detection probabilities are assumed to be the same for all targets. Hence, the PGFL of the predicted target-originated measurement process is 
\begin{align}
	\Psi_{\text{PHD}}^{\text{targets}}(h,g)=\,&G_{N}^{\text{PPP}}\big (\Psi_{\text{BMD}}(h,g) \big ),
\end{align}
where $\Psi_{\text{BMD}}(h,g)$ is given by (\ref{BayesMarkovMissedDet}). The target-originated measurement process is superposed with the independent clutter process, so the PGFL of the PHD filter is the product 
\begin{align}
	\label{pgflPHDOriginal}
	\Psi_{\text{PHD}}(h,g)=\,&\Psi_{C}^{\text{PPP}}(g)\,G_N^{\text{PPP}}\big(\Psi_{\text{BMD}}(h,g)\big).
\end{align}
Substituting (\ref{PGFLforClutter}), (\ref{pgfPoissonTargets}), and (\ref{BayesMarkovMissedDet}) gives the explicit form 
\begin{align}
	\label{pgflPHD}
	&\Psi_{\text{PHD}}(h,g)= \exp\Bigg( -\Lambda -\overline N + \Lambda \int_Y g(y) p_\Lambda(y) \, dy \nonumber\\ 
	&+\overline N\int_S h(x) \mu(x) \bigg(a(x)+b(x) \int_Y  g(y) p(y|x) \,dy\bigg)  dx\Bigg).
\end{align} 
\indent It has been noted (see, e.g. \cite{StoneStreitBMTT} and \cite{StreitDiscreteIntensity}) that the mathematical form of the PGFL of the Bayes posterior process -- before approximation to close the Bayesian recursion -- is the product of a PPP and $m$ Bernoulli target processes, where $m$ denotes the number of measurements.  In other words, the Bayes posterior process is the superposition of $m$ Bernoulli processes and a PPP. This form is closely related to that of the multi-Bernoulli filters discussed below in Section \ref{sec:multiBernoulliIntensityFilter}.\\
\indent In \cite{ChallaPHDJIPDA} the connection between a Gaussian Mixture implementation of the PHD intensity filter and the JIPDA filter is investigated. It is shown that under certain conditions (each target has a linear Gaussian dynamical model, target survival and detection probabilities are state independent, no explicit target birth and spawning event) the \textit{composite density} (of the JIPDA) conforms to the definition of probability hypothesis density (see \cite[Sec. 4]{ChallaPHDJIPDA}).

\subsection{CPHD Intensity Filter}
\label{sec:cphdIntensityFilter}
The CPHD intensity filter is proposed in \cite{Mahler07}, \cite{MahlerBook}. It propagates besides the intensity additionally the cardinality distribution and its PGFL. The assumptions made are essentially the same as in the PHD intensity filter. The differences are that the CPHD intensity filter propagates the cardinality distribution (\ref{cardinalityDistribution}), that the clutter process is given by an i.i.d. cluster process as defined in (\ref{pgflIIDClusterClutter}) and that the PGF of the number of present targets is also given by an i.i.d. cluster process, that is 
\begin{align}
	G_{N}^{\text{Cluster}}(z) = \sum_{n=0}^\infty p_N^{\text{Cluster}}(n)z^n,
\end{align}
where $p_N^{\text{Cluster}}(\cdot)$ is the distribution of the number of targets in $S$. The joint PGFL of the CPHD intensity filter is 
\begin{align}
	\label{PGFLCPHD}
	\Psi_{\text{CPHD}}(g,h) = \Psi_{C}^{\text{Cluster}}(g)\,G_{N}^{\text{Cluster}}(\Psi_{\text{BMD}}(g,h)).
\end{align}
The PGFL of the CPHD intensity filter reduces to the PGFL of the PHD intensity filter if the target and measurement processes are both given by PPPs.\\
\indent In \cite{Mallick} a CPHD filter is proposed that uses a fixed number of targets. Note that if this CPHD filter employs a clutter process that is given by a PPP it is equivalent to the JPDAS filter proposed in (\ref{PGFLforJPDAS}).\\
\indent The PGF $F^{\Xi|\Upsilon}(z)$ and the intensity $f^{\Xi|\Upsilon}(s)$ of the Bayes posterior process $\Xi|\Upsilon$ are defined as in (\ref{PGFNumberOfTargets}) and (\ref{intensity}) with respect to the joint PGFL $\Psi_{\text{CPHD}}(g,h)$. To close the Bayesian recursion, the posterior process is approximated by a point process $\widehat{\Xi|\Upsilon}$. The PGF of the number of targets in the approximating process is taken equal to that of the original process:
\begin{equation}
F^{\widehat{\Xi|\Upsilon}}(z)\equiv F^{\Xi|\Upsilon}(z).\nonumber
\end{equation} 
The probability distribution of  $\widehat{\Xi|\Upsilon}$ conditioned on $n$ targets is defined by the product approximation
\begin{align}
p_n^{\widehat{\Xi|\Upsilon}}(s_1,...,s_n) = \prod_{i=1}^{n}p^{\widehat{\Xi|\Upsilon}}(s_i),
\end{align} 
where the PDF for a single target is defined by 
\begin{align}
p^{\widehat{\Xi|\Upsilon}}(s)=f^{{\Xi|\Upsilon}}(s) \bigg{/} \frac{dF^{\Xi|\Upsilon}}{dx}(1).
\end{align}
In words, the approximate process $\widehat{\Xi|\Upsilon}$ is a cluster process whose PGF of target number is chosen to be equal to that of the posterior process, and whose points are i.i.d. distributed with PDF proportional to the normalized intensity of the posterior process. 

\subsection{Generalized PHD Intensity Filters}
\label{sec:generalizedPHDIntensityFilters}
The generalized PHD intensity filter was first proposed in \cite{ClarkGeneralizedPhd} and shares most of the assumptions of the standard PHD intensity filter presented in Section \ref{sec:phdIntensityFilter}. The predicted target process is a PPP and the updated target process is approximated by a PPP due to the same arguments as in Section \ref{sec:phdIntensityFilter}. The differences are the target-originated measurement process and the clutter model. In Section \ref{sec:phdIntensityFilter} it is assumed that one target generates at most one measurement per sensor-scan. Furthermore, it is assumed that the clutter model is given by (\ref{PGFLforClutter}), that is clutter is Poisson-distributed. The generalized PHD intensity filter relaxes these assumptions. First, the clutter model can be chosen arbitrary. Therefore, let $\Psi_C^{\text{gen}}(g)$ be the PGFL of a specified, but arbitrary, clutter process. Furthermore, let
\begin{align}
	\label{generalizedBayesMarkovMissedDetection}
	&\Psi_{\text{BMD}}^{\text{gen}}(h,g) \equiv %\int_S h(x) \mu(x) \Psi_{T|x}(g)  \,dx	\nonumber\\
	\int_S h(x) \mu(x) \bigg (p_n(\emptyset|x)+\nonumber\\
	&\sum_{n\ge 1} \frac 1 {n!} \int_{Y^n}  \prod_{i=1}^n g(y_i) p_n(y_1,...,y_n|x) \,dy_1...dy_n \bigg ) \, dx
\end{align} 
be the PGFL of the target measurement process, where $p_n(y_1,...,y_n|x)$ denotes the generalized symmetric likelihood function, which is defined on  $Y^n$. As mentioned in \cite{ClarkGeneralizedPhd} the probability of detection is defined more generally than for the standard PHD filter. Therefore, $p_n(\emptyset|x)$ denotes the probability of a missed detection and $p_n(y_1,...,y_n|x)$ includes the probability of detecting $(y_1,...,y_n)^T\in Y^n$ given a specific target state $x\in S$. The PGFL of the generalized PHD intensity filter is therefore given by
\begin{align}
	\label{pgflGeneralizedPHD}
	&\Psi_{\text{GenPHD}}(h,g)= \Psi_C^{\text{gen}}(g)\, G_N\big(\Psi_{\text{BMD}}^{\text{gen}}(h,g)\big),  
\end{align} 
which was first proposed in \cite{ClarkGeneralizedPhd}. 
%Can this PGFL can be used to derive Kastella's multitarget filter \cite{Kastella}. 
%and in case a Poisson-clutter model is chosen it is given by
%\begin{align}
%\label{pgflGeneralizedPoissonClutterPHD}
%&\Psi_{\text{GenPHD}}^{PPP}(h,g)= \exp\Bigg( -\Lambda -\overline N + \Lambda \int_Y g(y) p_\Lambda(y) \, dy +\nonumber\\ 
%&\overline N\int_S h(x) \mu(x) \Bigg(p^{\text{gen}}(\emptyset|x) + \nonumber\\
%&\sum_{n\ge 1} \frac 1 {n!} \int_{Y^n}  \prod_{i=1}^n g(y_i) p^{\text{gen}}(\{y_1,...,y_n\}|x) \,dy_1...dy_n \bigg )  dx\Bigg).
%\end{align}

\subsection{Multi-Bernoulli Intensity Filters}
\label{sec:multiBernoulliIntensityFilter}
\indent A Bernoulli random variable (trial) has two outcomes, or events, that are usually labeled ``success" and ``failure." Its PGF is $G^{\textnormal{Ber}}(z)\equiv 1-q+q z$,  $z\in\mathbb C$, where $q\in[0,1]$ is the probability of success. A multi-Bernoulli random variable is defined to be the number of successes in $n$ mutually independent Bernoulli trials. If $q_i\in[0,1]$ is the probability of success in the $i^{\textnormal{th}}$ trial, then the PGF of the number of successes is the product
\begin{align}
	G^{\text{multiBer}}(z)  \equiv \prod_{i=1}^n \left(1-q_i + q_i z\right).
\end{align}
Bernoulli models were used in Section \ref{sec:notation} for target detection modeling. They are used here to model target existence.\\
\indent The multi-Bernoulli intensity filter proposed in \cite{MahlerBook} makes the same assumptions as the PHD intensity filter, except that the prior target process is assumed to be given by a multi-Bernoulli process. Enhanced versions, implementations and numerical examples of the multi-Bernoulli intensity filter can be found in \cite{VoMB1}, \cite{VoMB2}.\\ 
\indent The predicted target process is again a multi-Bernoulli process. This is guaranteed by assuming the birth process to be a multi-Bernoulli process, which is independent of the target process. Target death is modeled by independent probabilistic thinning. Targets are assumed to be independent Markovian processes having the same motion model $p_0(x|x_0)$ described in Section \ref{sec:generalProbabilityModels}.\\ 
\indent The Bayes posterior process is not a multi-Bernoulli process. Thus, it is approximated by a multi-Bernoulli process consisting of a superposition of two independent multi-Bernoulli processes, one modeling data-induced targets, the other modeling existing targets. This assures the Bayesian recursion to be closed.\\
\indent The predicted target process is a multi-Bernoulli process with expected number of targets $\overline N \equiv E(N)$. Let $n \equiv \lfloor \overline N \rfloor$ be the largest integer less than or equal to $\overline N$. The PGFL of the multi-Bernoulli target process is given by 
\begin{align}
	\label{pgfBernoulliTargets}
	\Psi_{N{\text{MB}}}(h,g) = \prod_{i=1}^{n}\left(1-\chi_i+\chi_i\, \Psi_{\text{BMD}(i)}(h,g)\right).
\end{align}
Here, $\chi_i$ denotes probability that the $i^{\textnormal{th}}$ predicted (hypothesized) target is indeed a target, i.e., that it exists. Analogously to (\ref{pgflForJIPDADataDriven}), the PGFL for data-induced targets is given by
\begin{align}
	\Psi_{M{\text{MB}}}(h,g) = \prod_{j=1}^m\left(1-\gamma_j+\gamma_j\, \Psi_{\text{BMD}(j)}^{\text{Data}}(h,g)\right).
\end{align}
In general a cluster process as defined in (\ref{pgflIIDClusterClutter}) is proposed here to model clutter. Then, due to the superposition of the target-originated measurement process with the independent clutter process, the PGFL of the multi-Bernoulli filter is
\begin{align}
	\Psi_{\text{MB}}^{\text{Cluster}}(g,h)=\Psi_{C}^{\text{Cluster}} (g)\,\Psi_{N\text{MB}}(h,g)\,\Psi_{M{\text{MB}}}(h,g).\label{mB-PPPclutter}
\end{align}
Substituting a PPP clutter model for the more general cluster process model in (\ref{mB-PPPclutter}) gives 
\begin{align}
	\label{pgflMultiBernoulli}
	\Psi_{\text{MB}}(g,h)=\Psi_{C}^{\text{PPP}}(g)\,\Psi_{N\text{MB}}(h,g)\,\Psi_{M{\text{MB}}}(h,g).
\end{align} 
This is the PGFL of the multi-Bernoulli intensity filter given in \cite{MahlerBook}. \\ 
\indent Comparing the joint PGFL of the data-driven JIDPA filter from (\ref{pgflForJIPDADataDriven}) with the PGFL of the multi-Bernoulli filter in (\ref{pgflMultiBernoulli}), it is evident that the PGFLs of both filters differ only in the application of superposition. The data-driven JIPDA has as many target state spaces as there are targets and measurements. In contrast, the targets within the multi-Bernoulli filter all share the same target state space. Therefore, the multi-Bernoulli filter can be described as a data-driven JIDPA filter which employs superposition. The multi-Bernoulli filter is also described as a superposed single-scan MHT algorithm in Section \ref{JIPDAandMHT}. \\
\indent The connections between the multi-Bernoulli, the JIPDA, and the MHT filters are studied in terms of random finite sets in \cite{Williams}. The multi-Bernoulli filter derived in \cite{Williams2} is closely related to the set JPDA (SJPDA) \cite{setJPDA} filter.\\
\indent Various extensions of the multi-Bernoulli filter are proposed \cite{Reuter} including labeled versions to keep account of the target identity. The labeled multi-Bernoulli process is described in \cite{Reuter} as a random finite set on the Cartesian product of the state and label spaces, but this is inaccurate, for then the labels would be random, which they are not. As it can be seen from Table \ref{tableOfPointillistFilters} and the discussions above, the labeled version of the multi-Bernoulli filter corresponds to the JIPDA filter since the incorporation of target labels is equivalent to not superposing the targets onto one state space. 

%% file: hybridPointillistFilters.tex
The filters presented in this section assume at most $n$ target groups to be present. Therefore, each target group has its own state space and, within these groups, targets are superposed.\\
\indent It is also sometimes possible to superpose some targets and not others. Selective superposition has its uses. For example, targets that are well-separated could be superposed in one target state space, thereby possibly reducing computational complexity without incurring significant information loss, while the remaining targets are not superposed. Other examples involve highly mixed scenarios in which some groups of targets may have the same ``within-group" dynamical model (e.g., constant velocity), but with different groups having different dynamical models. Selective superposition  procedures are an additional decision step, which is outside the scope of this paper.

\subsection{Joint PHD Intensity Filter}
\label{sec:jointPHD}
The joint PHD intensity filter \cite{StreitJointPHD} assumes that exactly $n\ge 1$ target groups are present. The groups are mutually independent. Each group has its own state space $S_i$, $i=1,...,n$, and the targets in a group generate at most one measurement. Analogously to the PHD intensity filter (see Section \ref{sec:phdIntensityFilter}) the target groups constitute a finite point process at the recursion start time. The PGFL of the updated process is not necessarily a PPP and thus it is approximated by a PPP for closing the Bayesian recursion. The predicted target process is a PPP, since independent probabilistic thinning (target death) is assumed per target group and independent Markovian targets of one target group are assumed to have the same motion model.\\
\indent The clutter process is assumed to be a PPP. Furthermore, the processes of the target groups and clutter are assumed to be mutually independent and hence the PGFL of the joint PHD intensity filter is given by
\begin{align}
	\label{jointPHDPGFL}
	\Psi_{\text{JointPHD}}(h_1,...,h_n,g) = \Psi_{C}^{\text{PPP}}(g) \prod_{i=1}^n G_{N_i}^{\text{PPP}}\big (\Psi_{\text{BMD}}(h_i,g)\big),
\end{align}
where $G_{N_i}^{\text{PPP}}(z) = e^{-\overline N_i+\overline N_i z}$ is the PGF of the target number of the $i$-th target group and $\overline N_i$ is the expected number of targets of the predicted target process.

\subsection{Joint Generalized PHD Intensity Filter}
\label{sec:jointGeneralizedPHD}
Analogously to the joint PHD intensity filter, the joint generalized PHD filter assumes exactly $n\ge 1$ target groups to be present and these groups are mutually independent. As in the joint PHD intensity filter each group has its own state space $S_i$, $i=1,...,n$. Targets in a group are allowed to possess an arbitrary target-oriented measurement process, that is targets in a group can generate more than one measurement per sensor-scan. To the knowledge of the authors this filter has not been published yet. The target groups constitute a PPP at the recursion start time. Since the Bayes posterior process is not necessarily a PPP it is approximated by a PPP to close the Bayesian recursion. Due to the same arguments as in Section \ref{sec:jointPHD} the predicted target process is a PPP.\\
\indent As in Section \ref{sec:generalizedPHDIntensityFilters} the clutter model can be chosen arbitrarily. Thus, the joint PGFL of the joint generalized PHD intensity filter is given by
\begin{align}
	\label{pgflJointGeneralizedPHD}
	\Psi_{\text{JointGenPHD}}(h_1,...,h_n,g)=\Psi_C^{\text{gen}}(g)\prod_{i=1}^n G_{N_i}^{\text{PPP}}\big(\Psi_{\text{BMD}}^{\text{gen}}(h_i,g)\big),
\end{align}
where $\Psi_C^{\text{gen}}(g)$ is the PGFL of the arbitrarily specified clutter model used in (\ref{pgflGeneralizedPHD}) of Section \ref{sec:generalizedPHDIntensityFilters}.

%% file: closingBayesianRecursion.tex
\indent For all pointillist filters, the PGFL of the Bayes posterior point process can be characterized by a ratio of functional derivatives of the joint PGFL \cite{MahlerBook}, \cite{StreitPGFL}, \cite{ClarkGeneralizedPhd}. For discrete multivariate distributions, the same kind of derivative ratio for the PGF of the Bayes conditional distribution has long been known \cite[Equ. (34.48), p. 11]{JohnsonKotz}. \\
\indent A Bayesian recursion is said to be closed, or exact, if the probability distributions of the prior and posterior processes have the same mathematical forms. (There is a class of  filters, outside the scope of this paper, that is closed in this sense \cite{Daum1}, \cite{Daum2}.) For the filters whose PGFLs are given in this paper, however, it can be verified by inspection that the prior and posterior distributions take different forms. Consequently, to close the recursion the posterior point process must be approximated by a process having the same form as the prior. \\
\indent Methods for closing the Bayes recursion for pointillist filters that superpose targets have been discussed above. These methods make critical use of the summary statistics of the Bayes posterior process. One problem, already mentioned in Section \ref{sec:JPDAS}, arises when the Bayes posterior process is approximated by a simpler process, e.g., a PPP, to close the Bayesian recursion. The problem is that the targets are i.i.d. samples of the same PDF and, intuitively speaking, they are drawn with replacement. Thus, even if the PDF has as many "bumps" (local maxima) as there are targets, the target samples can \textit{with nonzero probability} all be drawn from the same bump. These samples should be drawn \textit{without} replacement so that distinct targets are sampled only once. PPPs therefore mis-model the Bayes posterior multitarget state point process. \\
\indent  Closing the Bayes recursion is theoretically easier for non-superposed filters because there are distinct state spaces for each possible target. When all targets are assumed to exist, it can be shown from the PGFL that the intensity function (or PDF, if the PGFL is multilinear, as discussed in Appendix \ref{LinearMomentsandPDFS}) is defined over the Cartesian product of target state spaces. The difficulty is that the high dimensionality makes it impractical to carry an estimate of the joint intensity. One traditional statistical approach to reducing the dimensionality of multivariate problems is to approximate the joint distributions as a product of the $n$ univariate marginal distributions \cite{Grim}, \cite{JohnsonKotz}. The product form is reasonable given that targets are mutually independent. This approach is taken by JPDA \cite{BarShalom2}, \cite{BarShalom3}. The Bayes recursion is closed because the prior is assumed to have the same product form.  \\
\indent Non-superposed pointillist filters encounter other issues when, as in JIPDA, there is \textnormal{at most} one target in each target state space, i.e., when a target may or may not exist. In this case marginalizing over all but one target and conditioning on its target existence enables a conditional target state PDF to be found, together with the probability of target existence \cite{Musicki3}. Cycling through the targets one at a time gives a discrete-continuous marginal distribution for each target. Using the product approximation for the joint distribution closes the Bayes recursion just as in JPDA. \\ 

%% file: targetStateEstimation.tex
\indent The minimum Bayes risk  estimate is defined for finite point processes as the realization of the process that minimizes the expected value of a specified cost function \cite{StreitPGFL}. The Bayes estimate yields an estimate of the number of targets and their states. In many problems the minimum Bayes risk estimate is equivalent to a maximum \textit{a posteriori} (MAP) estimate for an appropriately defined cost function; however, this is not one of those problems because realizations of finite point processes usually have different numbers of points. \\
\indent The Bayes estimate is impractical; instead, pseudo-Map estimates are found using the summary statistics of the Bayes posterior process. The estimated number of targets, $\hat n_{\text{pseudoMAP}}$, is the maximum of the (discrete) posterior distribution of the number of targets, as determined from the PGFL (see Appendix A, (\ref{cardinalityDistribution})-(\ref{PGFNumberOfTargets})). Subsequently, $\hat n_{\text{pseudoMAP}}$ target states estimates are obtained from the intensity function (or first moment, or PHD) of the Bayes posterior process. Pseudo-MAP surrogates are intuitively reasonable when they correspond to stable local peaks of the intensity function. The local peaks are often unstable \cite{MahlerBook}, \cite{ErdincWillettShalom}, so practical difficulties abound. When two or more targets coalesce into one intensity peak, it is not clear how to proceed. A nontrivial theoretical objection is that the intensity (number of targets per unit state space) is a summary statistic and not a full multitarget PDF, so the meaning of estimates obtained from it are of unknown statistical character. Efforts to overcome these problems have been investigated elsewhere \cite{improvedSMCPHD}, \cite{zhaoExtraction}.\\
\indent Tracking filters that do not use target superposition maintain separate state spaces for each target. This makes the target state estimation problem relatively straightforward, at least theoretically. As discussed in the previous section a common way to handle large numbers of targets is to write the joint posterior distribution as a product of marginal distributions. 

%% file: howToDesignATrackingFilter.tex
\begin{figure*}
	\center
	\includegraphics[width=1.0\textwidth]{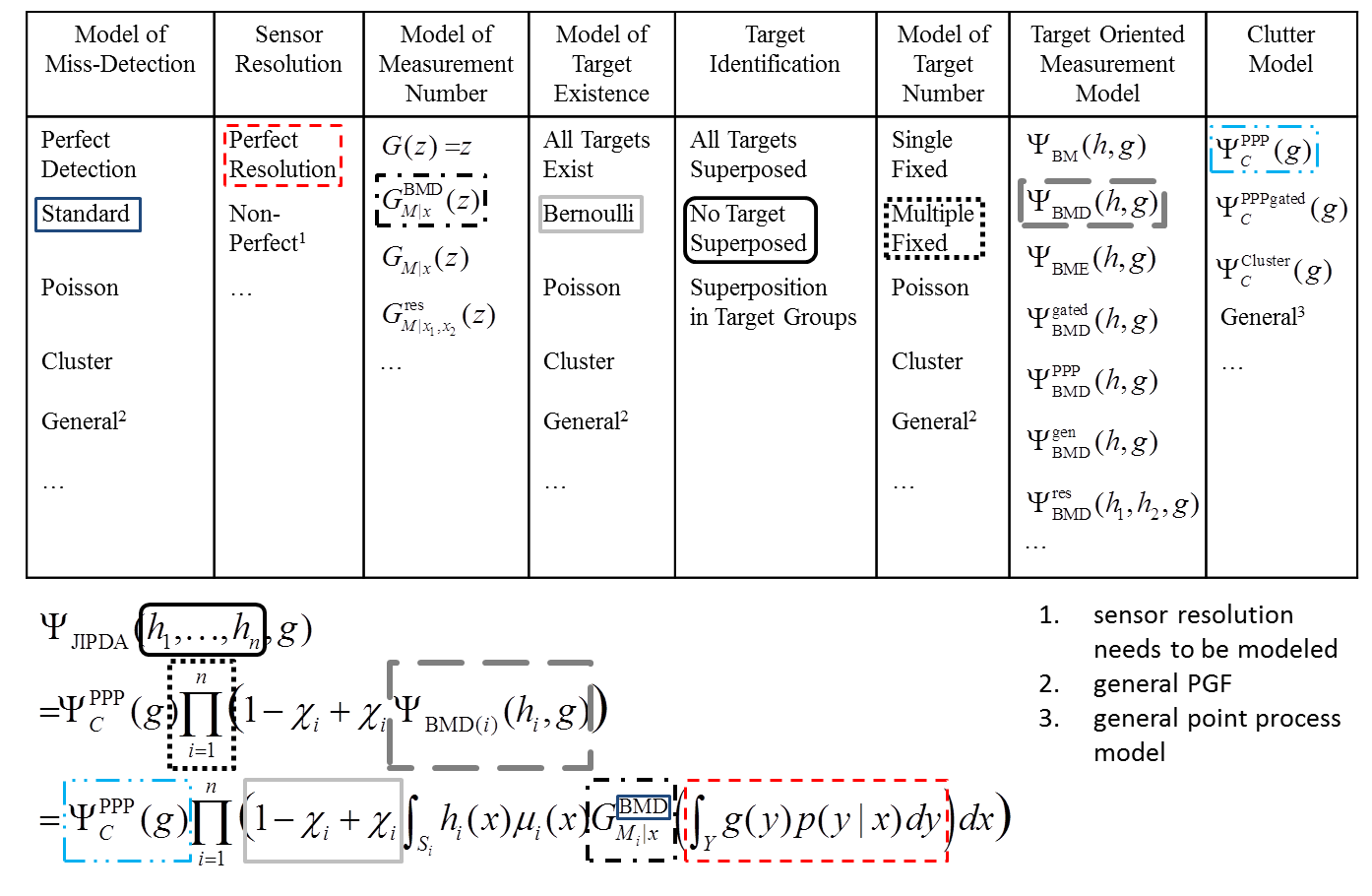}
	\caption{\textit{Discovery Step}.  A tracking filter is fully characterized by its joint target-measurement PGFL. This figure depicts the palette of available point process models for targets and measurement used in the PGFLs of the filters studied in sections up to and including Section \ref{sec:howToDesignATrackingFilter}. It highlights the construction of the PGFL for the JIPDA filter.}
	\label{fig:DiscoveryStep}
\end{figure*}
\begin{figure*}
	\center
	\includegraphics[width=1.0\textwidth]{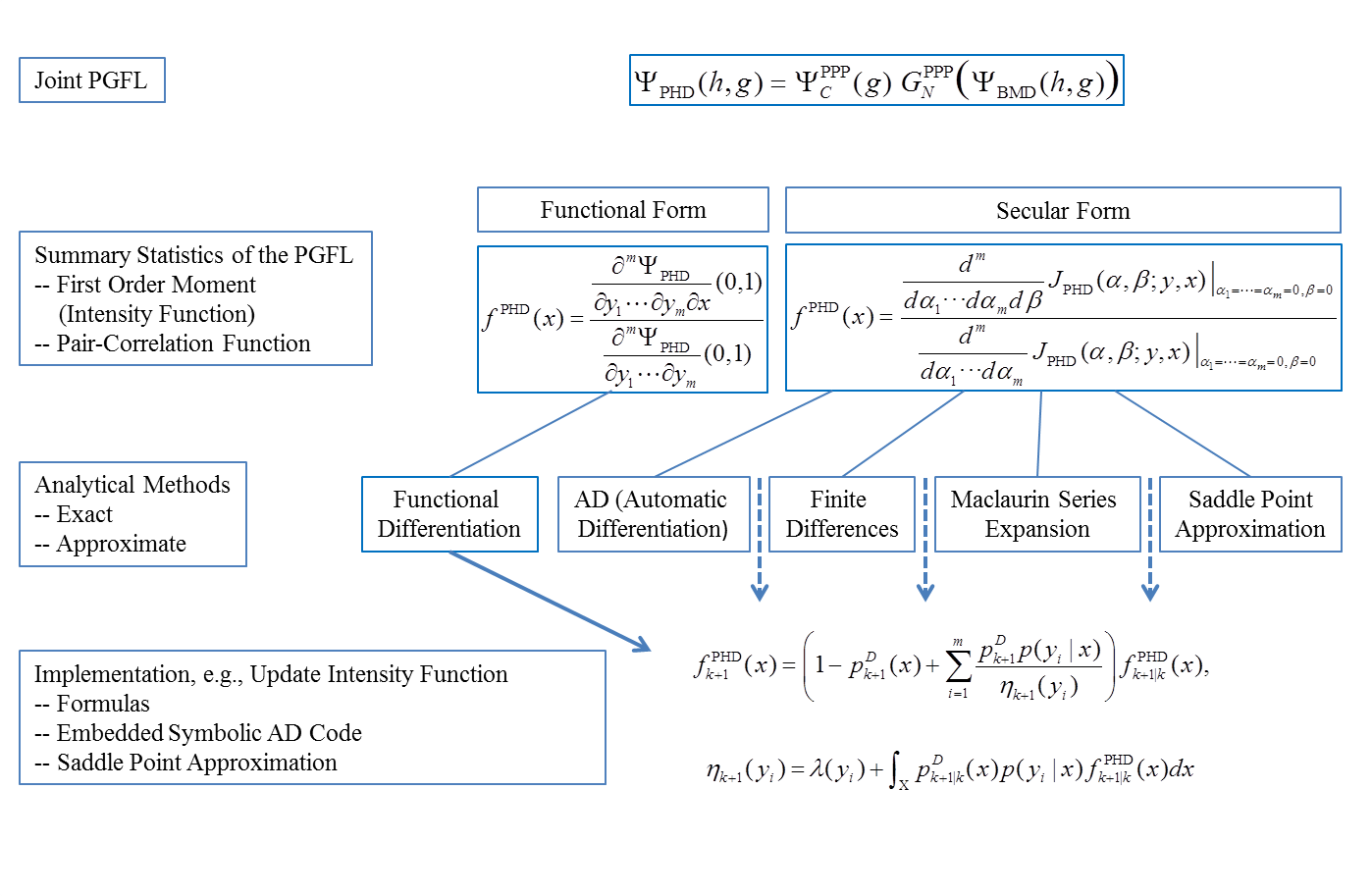}
	\caption{\textit{Analytical Step}. Derive summary statistics of the respective tracking filter by differentiating the PGFL of the Bayes posterior point process. The intensity function $f$ in the general case is given by (\ref{intensity}). This figure depicts the variety of choices available for any PGFL, while indicating those made for the PHD intensity filter. Symbolic functional derivatives of the PGFL must be done by hand, but lead to explicit formulas. The secular form of the PGFL (see (\ref{multivariateSecFun}) and (\ref{secularDerivative})) can be differentiated symbolically using widely available software. Exact numerical values  for particle weights can be found by AD. Derivatives of all orders of the secular PGFL can be written using the Cauchy integral method, which lends itself to saddle point approximation. See Section \ref{sec:alternativesToFunctionalDifferentiation}.}
	\label{fig:DerivationStep}
\end{figure*}

\indent The solution of a practical tracking problem using finite point processes can be separated into two methodologically different steps. First, the PGFL is constructed using different application-specific ingredients, such as the target, measurement and clutter model, complete, partial or no superposition of target states, the target-generated measurement model, the sensor resolution model, etc. This process is called the \textit{Discovery Step} in Section \ref{sec:background}, and its component steps are depicted in Figure \ref{fig:DiscoveryStep}. Up to this point, the paper has studied many different filters using this basic procedure. It can be seen from Figure \ref{fig:DiscoveryStep} that several combinations with various models for the ingredients of a tracking filter are possible, and this leads to an enormous number of different possible tracking filters. The enthusiastic tracking engineer can use Figure \ref{fig:DiscoveryStep} as an inspiration to design new customized tracking filters that fit the problem of current interest.\footnote{The situation calls to mind James Clerk Maxwell's famous words: ``\textit{There is nothing more practical than a good theory}."}\\
\indent If the tracking problem is modeled by constructing the PGFL, the second step is to derive the formulas needed for the implementation of the particular tracking filter, i.e., to compute the summary statistics (first or higher order moments, distribution of the number of targets, etc.) needed to close the Bayesian recursion. These statistics are given by ratios of functional derivatives with respect to the Dirac delta, evaluated at different points of the state and measurement spaces \cite{MahlerBook}, \cite{ClarkGeneralizedPhd}, \cite{StreitPGFL} (see also Appendix \ref{sec:AppendixA}). Figure \ref{fig:DerivationStep} visualizes this \textit{Analytical Step} using the PHD filter \cite{MahlerBook}, \cite{Mahler03},\cite{Mahler07}, the first filter to be derived from a PGFL.\\
\indent Alternatives to functional differentiation are discussed in Section \ref{Alternatives to Functional Differentiation}. Several of the methods presented there make it possible to  assess particle filter performance on simulated data sets quickly and reliably, that is, they provide a very cost effective way to explore the design space without the need for expensive hand-crafted code.\\
\indent The framework of PGFLs and point processes used in the previous sections is of more than purely theoretical interest. First, it shows similarities and differences between existing well-known tracking filters. For example, the consideration of how the superposition of targets is modeled within the framework of point processes brings to light the close connection of the multi-Bernoulli, the un-pruned/merged MHT and the JIPDA filter. Furthermore, it enables the reader to understand better the kinds of challenges that arise if new assumptions are made (e.g., on target superposition, Section \ref{sec:commentsOnTargetSuperposition}) or if old assumptions are altered---seemingly small changes can have disproportionate impact. The proposed design procedure therefore helps the experienced tracking engineer to understand existing filters and their connections. \\
\indent Second, the framework gives birth to an entirely new class (to the knowledge of the authors) of tracking filters, called hybrid pointillist filters, by a straightforward application of the assumption that target states are superposed only within a specific number of target groups. It is evident to ask for detailed numerical evaluations of this new class of filters; however, a close investigation of practical applications is not part of this work and will be presented in future publications.\\
\indent In addition to the important discoveries mentioned above, a practical working tracking engineer may still ask: What's in it for me? The answer is: Unifying tracking filters in a common framework offers the possibility of customized designs for application-specific tracking problems that engineers are confronted with in practical work. Once understood it offers an easy way of finding out which existing methods can be re-used and which parts of the problem have to be modeled in a different way. Thereby, the ingredients such as clutter model, target-generated measurement model, etc. of the existing tracking filters can be mixed in a large variety of combinations, as depicted in Figure \ref{fig:DiscoveryStep}. In the following, a guide to how to design a customized tracking filter is presented for a specific tracking scenario by making use of the essential cornerstone-parameters of the problem.\\
\indent Assume exactly two unresolved targets $x_1\in S_1,\ x_2\in S_2$ with probability of detection $p_{D,1}(x_1,x_2)$ and $p_{D,2}(x_1,x_2)$, respectively, to be present in a cluttered environment. The target's identity is of particular interest and one target generates at most one measurement. This problem has already been formulated in \cite{SvenssonUlmkeDanielsson}) (see also \cite{Blom}, \cite{Chang}) and should demonstrate how a tracking filter is formulated in terms of its joint PGFL. The following basic questions for designing a tracking filter are answered for each of the columns of Figure \ref{fig:DiscoveryStep} for the unresolved target problem. By following these questions, the columns of Figure \ref{fig:DiscoveryStep} provides a guide for formulating custom filters for other problems.

\begin{enumerate}	
	\item \textit{What is the model of target missed detections?}\\
	Choose a standard missed detection model. For ease of discussion it is assumed that $p_{D,j}(x_1,x_2)\equiv p_{D,j}(x_j)$, $j=1,2$, that is, the probability of detecting target $j$ depends only on target $j$. Adopting the notation (\ref{detectionprobs}) gives  $a_j(x_j)=1-p_{D,j}(x_j)$ and $b_j(x_j)=p_{D,j}(x_j)$, $j=1,2$.   
	\item \textit{How is the sensor resolution modeled?}\\
	The resolution of the targets depends on how close they appear in the sensor. For example, the sensor resolution function can be Gaussian, i.e.,
	\begin{align}
	f^{\textnormal{res}}(x_1,&x_2) \label{resFunction} \\
	=\,&e^{-\big(H_1(x_1) - H_2(x_2)\big)^T \Sigma^{-1} \big(H_1(x_1) - H_2(x_2)\big)/2},\nonumber
	\end{align}
	where the matrix $\Sigma$ is determined by the nature of the sensor and the system functions $H_1: S_1\to Y$, $H_2: S_2\to Y$ map the target states to points in the measurement space. For this choice, targets at $x_1\in S_2,x_2\in S_2$ are poorly resolved by the sensor if $f^{\textnormal{res}}(x_1,x_2) \approx 1$ and well resolved if $f^{\textnormal{res}}(x_1,x_2) \approx  0$. Whether or not unresolved target states are close depends on the system functions $H_1$ and $H_2$. The Gaussian function $f^{\textnormal{res}}$ can be replace by any reasonable function $f^{\textnormal{res}}: S_	1\times S_2\to[0,1]$ 
	provided that $f^{\textnormal{res}}(x_1,x_2)=0$ for all $x_1\in  S_1$, $x_2\in S_2$ with $H_1(x_1)=H_2(x_2)$. The resolution function, however defined, is used in the PGF of the number of measurements.
	
	\item \textit{How to model the number of measurements?}\\
	Let the PGF of the number of measurements be given by
	\begin{align}G_{M|x_1,x_2}^{\text{res}}(z) \equiv c_0 + c_1 z + c_2 z^2,
	\end{align}
	where
	\begin{align}
	c_0 \equiv  &\,a_1(x_1)a_2(x_2), \\
	c_1\equiv\,& a_1(x_1)b_2(x_2)+b_1(x_1)a_2(x_2) \nonumber\\
	&+ \,f^{\textnormal{res}}(x_1,x_2) b_1(x_1)b_2(x_2),\\
	c2\equiv & \,b_1(x_1)b_2(x_2) \big(1 - f^{\textnormal{res}}(x_1,x_2)\big)
	\end{align} 
	and $f^{\textnormal{res}}(\cdot,\cdot)$ is defined, for example, by (\ref{resFunction}). The number of target measurements depends on the distance between the two targets in the sensor space. If $x_1\approx x_2$, then $f^{\textnormal{res}}(x_1,x_2) \approx 1$ and the PGF of the number of targets is  
	\begin{align}
	G_{M|x_1,x_2}^{\text{res}}(z) \approx c_0 + c_1 z,\,\,\,\,\,z\in\mathbb C,	
	\end{align}
	which means that two poorly resolved targets yield at most one measurement with high probability. On the other hand, the PGF for well resolved targets is quadratic and, moreover, factors into two linear terms: 
	\begin{align}
	\label{factorizationJPDA}
		G_{M|x_1,x_2}^{\text{res}}&(z) = \, c_0 + c_1 z+c_2 z^2 \nonumber\\
		\approx & \,\big(a_1(x_1) + b_1(x_1) z\big)\big(a_2(x_2) + b_2(x_2) z\big)\nonumber\\
		= &\, G_{M|x_1}^{\text{BMD}}(z)\, G_{M|x_2}^{\text{BMD}} (z).
	\end{align} 
		Thus, well resolved detected targets yield two conditionally independent measurements, as expected.
	\item \textit{How is target existence model defined?}\\
	Two targets are assumed present, exactly as in JPDA, so no modeling of the target existence is needed.
	
	\item \textit{Is target identification needed?}\\
	Yes.  Therefore the joint PGFL will have two test functions, one for each target.
	
	\item \textit{How is the target number modeled?}\\
	A fixed number of targets is assumed.
		
	\item \textit{What is the target-generated measurement model?}\\
	It is assumed that a target generates at most one measurement per sensor scan. (This assumption can be relaxed; see  (\ref{BayesMarkovMissedDet}) and the extended version (\ref{generalizedBayesMarkovMissedDetection}).) Measurements, however, depend jointly on the states of both targets. The target-generated measurement model can be shown to be given by a modification of (\ref{BayesMarkovMissedDet}). It is defined by 
	\begin{align}
		\Psi_{\text{BMD}}^{\text{res}}&(h_1,h_2,g)  \nonumber\\
		\equiv&\,\int_{ S_1} \int_{ S_2} h_1(x_1) \mu_1(x_1)\, h_2(x_2) \mu_2(x_2)\nonumber\\
		\label{resPGFL}
		\times& \, G_{M|x_1,x_2}^{\text{res}}\left(\int_Y g(y) p(y|x_1,x_2)\, dy\right)\,dx_1dx_2,
	\end{align} 
	where $p(y|x_1,x_2)$ is an arbitrarily specified likelihood function that is conditioned on both targets. It is worth pointing out that the PGFL (\ref{resPGFL}) reduces to the PGFL of the standard JPDA filter for two targets under two assumptions: ($i$) the factorization (\ref{factorizationJPDA}) holds, that is, the targets are always well-separated; and ($ii$) the likelihood function $p(y|x_1,x_2)$ depends on only one target state not two. In this case $p(y|x_i)$ is used within $G_{M|x_i}^{\text{BMD}}(\cdot)$, $i=1,2$.

	\item \textit{Which clutter model is used?}\\
	In this case the clutter model is an arbitrary finite point process. We choose the Poisson clutter model $\Psi_{C}^{\text{PPP}}(g)$ and assume further that the clutter process is mutually independent of both target-generated measurement processes.
	
\end{enumerate}

\indent  The target-generated measurement and clutter processes are  mutually independent, so the joint PGFL is given by the product of their PGFLs:
\begin{align}
	\Psi_{\text{JPDA}}^{\text{res}}(h_1,h_2,g) = \Psi_{C}^{\text{PPP}}(g)\,\Psi_{\text{BMD}}^{\text{res}}(h_1,h_2,g).
\end{align}
This PGFL fully characterizes the two target example problem for unresolved measurements. This completes the tracking filter \textit{Discovery Step}.\\
\indent Sensor resolution issues increase the complexity of the tracking problem. In the language of PGFLs, this can be traced to the fact that the double integral involved in the target-generated measurement model $\Psi_{\text{BMD}}^{\text{res}}$ does not factor into a product of two integrals as in the perfectly resolved case, since the PGF of the number of measurements is not the product of first degree polynomials (cf. (\ref{resPGFL}). This has a significant impact in practice because computing double integrals is computationally much more  demanding than computing the product of two lower dimensional integrals. \\
\indent Secular functions can be defined  and derived for PGFLs with general target-generated measurement models. This can be seen directly for the unresolved target example using a method that is similar to that used in \cite{StreitSecular}. The proof of the result in the general case is more involved and will be presented elsewhere \cite{Paper1}. \\ 
\indent Another example of multiple integrals that do not factor is given by $\Psi_{\text{GenPHD}}(h,g)$ of the generalized PHD intensity filter (\ref{pgflGeneralizedPHD}). The reason in this case is that the generalized PHD filter enables a target to create more than one measurement per sensor scan and thus the integrals with respect to the measurement space do not factor.\\
\indent It is seen from the unresolved target example discussed above that in practice engineers can use the pointillist filter point of view to quickly design and characterize tracking filters for modeling specific problems of interest. \\
\indent The next step is the \textit{Analytical Step}, which is needed to obtain the explicit formulas for the implementation of the designed tracking filters. In the following section and Appendix \ref{sec:AppendixA} it is shown how summary statistics such as the first order moment (intensity function) and higher order moments (for pair-correlation) of the Bayes posterior process can be derived from the joint target-measurement PGFL found in the \textit{Discovery Step}. These statistics are used in filter implementations to close the Bayesian recursion. 

%% file: alternativesToFunctionalDifferentiation.tex
\label{sec:alternativesToFunctionalDifferentiation}
The distribution of the Bayes posterior point process is given by a ratio of functional derivatives with respect to the Dirac delta of the joint PGFLs presented in Sections \ref{sec:pointillistFiltersWithoutSuperposition} -- \ref{sec:hybridPointillistFilters}. Functional differentiation is needed to derive the summary statistics that are used to define the respective tracking filters and close the Bayesian recursion. \\
\indent This section presents alternatives to functional differentiation. Secular functions, which were first presented in \cite{StreitSecular}, are defined and methods for differentiating them are presented. Further details can be found in \cite{StreitSecular}, \cite{Fusion2015Streit}, and \cite{Paper1}. 

\subsection{Secular Functions}
\label{sec:defineSecularFns}
In the following, the definition of a secular function from \cite{StreitSecular} is presented for univariate and multivariate PGFLs. Details are omitted and can be found in \cite{StreitSecular}, \cite{Paper1}. \\
\indent Let $\Psi\in\mathcal P_2$, where $\mathcal P_2$ is defined in (\ref{PGFLsOfInterest}). The PGFLs of all investigated PGFLs in this work fall into this class of PGFLs and thus all tracking filters presented here can be derived by the following methods. For the univariate case the secular function is defined by 
\begin{align}
	\label{secularUnivariate}
	J(\alpha;c)\equiv &\underset{\lambda\searrow 0}{\lim}\Psi(h+\alpha\gamma_\lambda^c),
\end{align}
where $x\in \mathcal X\subseteq \mathbb R^d$, $\alpha\in \mathbb C^m$, $c\in\mathbb R^m$, $\lambda>0$, $h:\mathcal X\to\mathbb R$. Here, $\{\gamma^{c_i}_\lambda\}_{\lambda>0}$ is a family of test functions for the Dirac delta. For multivariate PGFLs the secular function is defined by
\begin{align}
	\label{secularMultivariate}
	J( \alpha;x) \equiv& \underset{\lambda\searrow 0}{\lim}\Psi\left(h(x) + \sum_{i=1}^m\alpha_i\gamma_\lambda^{c_i}(x)\right)\\
	\label{secular1Notation}
	= &\Psi\left(h(x) + \sum_{i=1}^m\alpha_i\delta^{c_i}(x)\right),
\end{align}
where (\ref{secular1Notation}) uses the Dirac delta as an evaluation operator, not a function. The definition (\ref{secularMultivariate}) can be interpreted as the limit of an approximation to the Dirac delta \cite{Alt}, \cite{EinfDistr}. Further discussion of the definitions (\ref{secularUnivariate}) and (\ref{secularMultivariate}) can be found in \cite{Paper1} and \cite{StreitSecular}. For $x\in\mathcal X$ and $\alpha\in\mathbb C^m$, it holds that
\begin{align}
	\frac{\partial \Psi}{\partial x}(h) = \frac{d J}{d\alpha}(0;x)\equiv  \frac{d}{d\alpha}J(\alpha;x)\Big|_{\alpha=0},
\end{align}
and for  $ x = \{x_1,\dots,x_m\}$ , $x_i\in\mathcal X$, $i=1,\dots,m$,
\begin{align}
	\label{SecularFunctionsMultivariate}
	\Psi_{x}(h)\equiv\,&\frac{\partial^m\Psi(h)}{\partial x_1\cdots\partial x_m}  \nonumber\\
	= &\, \frac{d^m}{d\alpha_1\cdot\cdot\cdot \alpha_m}J( \alpha;x)\Big|_{\alpha_1=\cdots=\alpha_m=0} \equiv J_\alpha (0;x),
\end{align}
which shows that the G\^{a}teaux derivative with respect to multiple Dirac deltas is identical to the first order mixed derivative of the corresponding secular function.\\
\indent Secular functions can also extended to joint PGFLs (see \cite[Sec. 4.2]{StreitSecular}). Let $\Psi(g,h)$ be the joint PGFL of the Bayes posterior process, where $g$ and $h$ are bounded functions on $Y\subseteq\mathbb R^{d_2}$ and $S\subseteq \mathbb R^{d_1}$, $d_1,d_2>0$. Then, the secular function is given by
\begin{align}
	\label{multivariateSecFun}
 &J( \alpha, \beta; y, x) \nonumber \\ 
	&\equiv\underset{\lambda_1\searrow 0}{\lim} \underset{\lambda_2\searrow 0}{\lim} \Psi\left(\sum_{i=1}^m\alpha_i\gamma_{\lambda_1}^{y_i}(y),1+ \sum_{j=1}^n\beta_j\gamma_{\lambda_2}^{x_j}(x)\right)\\
	\label{secular2Notation}
	&= \Psi\left(\sum_{i=1}^m\alpha_i\delta^{y_i}(y),1+ \sum_{j=1}^n\beta_j\delta^{x_j}(x) \right),
\end{align}
where (\ref{secular2Notation}) uses the Dirac delta as an evaluation operator in the same way as it was used in (\ref{secular1Notation}). It holds that
\begin{align}
	\label{secularDerivative}
	J_{\alpha \beta}(\alpha,\beta;y,x)\Big{|}_{\alpha=0,\beta=0}=\Psi_{y x }(g,h)\Big{|}_{g(\cdot)=0,\ h(\cdot)=1}.
\end{align}
\indent The $n^{\textnormal{th}}$ factorial moment (\ref{nthFactorialMoment}) of the Bayes posterior point process can be computed using secular functions by
\begin{align}
	m_{[n]}(x_1,\dots,x_n|y)\, %&\Psi_{x}(y|h)\Big{|}_{h(\cdot)=1} \nonumber\\
	%\equiv\,& \frac{\partial^n}{\partial x_1\cdots\, \partial x_n} \Psi(y,h)\Big{|}_{h(\cdot)=1}\nonumber\\
	=\,& \frac{\Psi_{ y  x}(0,1)}{\Psi_{ y}(0,1)},
\end{align}
where 
\begin{align}
	\Psi_{ y  x}(0,1) \equiv \frac{\partial^m}{\partial y_1\cdot\cdot\cdot \partial y_m}  \frac{\partial^n}{\partial x_1\cdot\cdot\cdot \partial x_n} \Psi(g,h)
\end{align}
and 
\begin{align}
	\Psi_{ y}(0,h)\equiv \frac{\partial^m}{\partial y_1\cdot\cdot\cdot \partial y_m}\Psi(g,h)\Big{|}_{g(\cdot)=0}.
\end{align}

\subsection{Differentiating Secular Functions}
The following sections present methods for differentiating secular functions.

\subsubsection{Application of Cauchy's Residue Theorem -- Saddle Point Methods}
Consider the complex function
\begin{align}
\alpha\mapsto J( \alpha;x) \equiv  \underset{\lambda\searrow 0}{\lim}\Psi\left(h(x) + \sum_{i=1}^m\alpha_i\gamma_\lambda^{c_i}(x)\right),
\end{align}
where $x\in \mathbb R^n$, $n\ge 0$, $\alpha\in \mathbb C^m$, $\Psi \in \mathcal P_2$ and $h\in\mathcal H$. Then, $J(\cdot;x)$ is analytic in some open region of the complex plane $\Omega\subset\mathbb C^m$ containing $(0,\dots,0)\in\mathbb C^m$ \cite[Sec. 4]{Moyal}. Denote by $\Omega_1,\dots,\Omega_m\subset \mathbb C$ the projections of the open region $\Omega$ to from $\mathbb C^m$ to $\mathbb C$.  Thus, Cauchy$'$s coefficient formula \cite[p.236]{Flajolet} can be applied to $\alpha\mapsto J(\alpha;x)$, $x\in\mathbb R^n$. Hence, the ordinary derivative of $J(\cdot;x)$ is given by
\begin{align}
\label{CauchyResiduum}
&\frac{d^m}{d\alpha_1\cdot\cdot\cdot d\alpha_m} J(\alpha;x)\Big|_{\alpha_1=\cdot\cdot\cdot=\alpha_m=0}\nonumber\\
& = \!\frac{1}{(2\pi i)^m}\! \int_{C_1}\!\!\!\dots\!\int_{C_m}\!\frac{J(\zeta;x)}{(\alpha_1-\zeta_1)^2\cdots(\alpha_m-\zeta_m)^2}\,d\zeta_1\cdots d\zeta_m,
\end{align} 
where $C_1,\dots,C_m$ in $\mathbb C$ are simple loops of $\Omega_1,\dots,\Omega_m$ encircling $0\in\mathbb C$, respectively. Due to the concept of secular functions and Cauchy's coefficient theorem (which is an implication of Cauchy's residue theorem), the first factorial moment can be determined by contour integration. Furthermore, note that the numerical approximation of contour integral with powerful saddle point methods has been extensively studied (see \cite{Bleistein1}, \cite{Flajolet}) and can be used to evaluate (\ref{CauchyResiduum}) numerically. In \cite{Fusion2015Streit} saddle point approximations for the JPDA filter are investigated for SMC implementations.   

\subsubsection{Automatic Differentiation}
Automatic differentiation (AD) methods are presented in \cite{Griewank}. These techniques compute numerical values of derivatives of a function without finding the symbolic derivative. AD methods are not approximations but exact (up to machine precision). Since the (ordinary) derivatives of secular functions are identical to the functional derivatives of PGFLs, the method of AD can be used for evaluating the first factorial moment at any given point/particle, e.g., to determine exact particle weights. 

\subsubsection{Classical Finite Differences and Maclaurin Series Expansion}
Since the first factorial moment in \cite[Sec. 5]{StreitSecular} is given by the ratio of ordinary derivatives evaluated at zero, it can be determined using classical approximation techniques like classical finite differences or a Maclaurin series expansion. The details on these methods can be found in \cite{StreitSecular}.

%% file: conclusionsAndFutureWork.tex
\label{Section::ConclusionsAndFutureWork}

\newcommand\Tstrut{\rule{0pt}{4.7ex}}         % = `top' strut
\newcommand\Bstrut{\rule[-3.0ex]{0pt}{0pt}}   % = `bottom' strut

\indent The ontological perspective of analytic combinatorics is new. It enables the relationships between known and new filters to be described succinctly in Table \ref{tableOfPointillistFilters}. As can be seen from the table, the unified analytic combinatoric perspective has significant added value -- it shows that target superposition is a highly flexible tool that can be used in many ways. The discovery of the new family of hybrid pointillist filters demonstrates that the new perspective is much more than a framework for comparing filter combinatorics. \\
\indent The pointillist family of tracking filters are those that can be derived via the joint PGFLs of target states and measurements, when the states and measurements are modeled by finite point processes. Three classes of pointillist filters are proposed. The best known class comprises filters that do not employ target superposition, that is, each target has its own state space. Such filters include the classical Bayes-Markov filter as well as the PDA, JPDA, IPDA, JIPDA, PMHT, and MHT filters. \\
\indent A newer class of pointillist filters are those that use target superposition, that is, all targets share the same state space. Examples of filters in this class are the PHD intensity, the CPHD intensity, the generalized PHD intensity, and the multi-Bernoulli filters. It is shown that some of these filters (multi-Bernoulli, CPHD with fixed number of targets) can be formulated and derived from superposed versions of the joint PGFLs of the non-superposed JPDA, JIPDA, and MHT tracking filters. \\
\indent Between these two main classes of filters is the class of hybrid pointillist filters, in which some of the targets are superposed and others not. Examples of this hybrid class presented in this paper are the joint PHD intensity and the generalized joint PHD intensity filters. \\
\indent An  alternative to functional differentiation of the PGFL is presented. This technique replaces the PGFL with an ordinary, or secular, function and functional derivatives by ordinary derivatives. The two methods are  equivalent theoretically; however, secular functions can be differentiated in many ways. Several are proposed in this paper, including a novel method involving the Cauchy residue theorem of classical complex analysis.\\
\indent Future work will investigate the multi-scan versions of the single sensor pointillist filters discussed in this paper, including the multi-scan MHT, as well as multi-sensor versions of these filters. Some of these PGFLs have already been derived and will be reported elsewhere.\\ 

\begin{table*} 
	\centering
	\scriptsize	
	\begin{tabular}{|l|l|l|l|l|l|l|l|l|l|}	
		\hline
		\label{tableOfPointillistFilters}
		\bf{Filter Name} & \bf{PGFL Notation} & \bf{Joint Target-Measurement PGFL} & \bf{Ref.} & \bf{Target}  & \bf{Clutter} & \bf{Missed} &  \bf{Super-} \Tstrut\\ 
		\ &  \ & \ & \bf{Eqn.} & \bf{Model} & \  & \bf{Det.} &  \bf{pos.} \Bstrut  \\  \hline		
		\bf{Bayes-Markov} & $\Psi_{\text{BM}}(h,g)$ & $\int_S \int_Y h(x) g(y) \mu(x) p(y|x) \,dy\,dx$ &(\ref{BayesMarkov})  & $ =1$ pt. tgt. & No & No & No \Tstrut \Bstrut \\
		\,\,\,\textit{Missed Detections} & $\Psi_{\text{BMD}}(h,g)$ & $\int_S h(x) \mu(x)\, G_{M|x}^{\text{BMD}}\big( \int_Y  g(y) p(y|x) \,dy\big) \, dx$ &(\ref{BayesMarkovMissedDet}) & $=1$ pt. tgt. & No & Std.  & No   \Bstrut\\	
		\,\,\,\textit{Missed Detections} & $\Psi_{\text{BME}}(h,g)$ & $\int_S h(x) \mu(x)\, G_{M|x}\big( \int_Y  g(y) p(y|x) \,dy\big) \, dx$ & (\ref{extendedBayesianTarget}) & $=1$ ext. tgt. & No & Gen.$^5$ & No  \Bstrut \\ 
		\,\,\,\,\,\,\textit{and Extended Target} &  \ & \ & \ & \ & \  & \ & \   \Bstrut \\ \hline
		\bf{PMHT} &  \ & \ & \ & \ & \ & \  & \  \Tstrut \Bstrut \\
		\,\,\,\textit{Without Superposition} & $\Psi_{\text{PMHT}}(h_1,\dots,h_n,g)$ & $\prod_{i=1}^{n}\Psi_{\text{BMD}(i)}^{\text{PPP}}(h_i,g)$  & (\ref{PGFLforPMHT}) & $=n$ ext. tgts. & No$^1$  & No$^8$  & No \Bstrut \\
		\,\,\,\textit{With Superposition} & $\Psi_{\text{PMHTS}}(h,g)$ & $\prod_{i=1}^{n}\Psi_{\text{BMD}(i)}^{\text{PPP}}(h,g)$  & (\ref{SuperposedPMHT}) & $=n$ ext. tgts. & No$^1$  & No$^8$ & Yes \Bstrut \\ \hline
		\bf{PDA} &  \ & \ & \ & \ & \ &\  & \ \Tstrut \Bstrut \\
		\,\,\,\textit{Without Gating} &  $\Psi_{\text{PDA}}^{\text{noGate}}(h,g)$ & $\Psi_{C}^{\text{PPP}}(g)\,\Psi_{\text{BMD}}(h,g)$&  (\ref{PDAnoGating}) & $=1$  pt. tgt. & PPP  & Std.  & No  \Bstrut \\
		\,\,\,\textit{With Gating} & $\Psi_{\text{PDA}}(h,g)$ & $\Psi_{C}^{\text{PPPgated}}(g)\,\Psi_{\text{BMD}}^{\text{gated}}(h,g)$ &  (\ref{PDAwithGating}) & $=1$  pt. tgt. & PPP  & Std. & No  \Bstrut	\\
		\,\,\,\textit{Extended Target} & $\Psi_{\text{PDAE}}(h,g)$ & $\Psi_{C}^{\text{PPP}}(g)\,\Psi_{\text{BME}}(h,g)$ & (\ref{PDAwithExtendedTarget}) &	$=1$  ext. tgt. & PPP & Gen.$^5$  & No \Bstrut \\\hline
		\bf{JPDA} &  \ & \ & \ & \ & \  & \ & \ \Tstrut \Bstrut \\
		\,\,\,\textit{Without Superposition} & $\Psi_{\text{JPDA}}(h_1,\dots,h_n,g)$ & $\Psi_{C}^{\text{PPP}}(g)\,\prod_{i=1}^n\Psi_{\text{BMD}(i)}(h_i,g)$  & (\ref{PGFLforJPDA}) & $=n$ pt. tgts. & PPP & Std.  & No \Bstrut \\ 
		\,\,\,\textit{With Superposition} & $\Psi_{\text{JPDAS}}(h,g)$ & $\Psi_{\text{JPDA}}(h,\dots,h,g)$  &  (\ref{PGFLforJPDAS}) & $=n$ pt. tgts. & PPP & Std.  & Yes  \Bstrut \\ \hline		
		\bf{IPDA} & $\Psi_{\text{IPDA}(h,g)}$ & $\Psi_{C}^{\text{PPP}}(g)\,\Big( 1-\chi+\chi\,\Psi_{\text{BMD}}(h,g) \Big)$  & (\ref{IPDA}) & $\le 1$ pt. tgt. & PPP & Std. & No \Tstrut \Bstrut \\ \hline
		\bf{JIPDA} &  \ & \ & \ &  \ & \ & \  & \  \Tstrut \Bstrut \\
		\,\,\,\textit{Without Gating} & $\Psi_{\text{JIPDA}}(h_1,\dots,h_n,g)$ & $\Psi_{C}^{\text{PPP}}(g)\,\prod_{i=1}^n\Big  (1-\chi_i+\chi_i\,\Psi_{\text{BMD}(i)}(h_i,g)\Big)$ & (\ref{pgflForJIPDAprocess}) & $\le n$ pt. tgts. & PPP & Std. & No \Bstrut \\	
		%\ & \ & $\times \, \Psi_{C}^{\text{PPP}}(g)$  &   \ & \ & \  &  \ & \ \Bstrut \\
		\,\,\,\textit{With Gating} & $\Psi_{\text{JIPDA}}^{\text{gated}}(h_1,\dots,h_n,g)$ & $\Psi_{C}^{\text{PPPgated}}(g)\,\prod_{i=1}^n\Big  (1-\chi_i+\chi_i\,\Psi_{\text{BMD}(i)}^{\text{gated}}(h_i,g)\Big)$ &  (\ref{pgflForJIPDAprocessGated}) & $\le n$ pt.. tgts. & PPP & Std.  & No \Bstrut \\	
		%\ & \ & $\times \, \Psi_{C}^{\text{PPPgated}}(g)$  &   \ & \ & \  &  \ & \ \Bstrut \\	
		\,\,\,\textit{With Superposition} & $\Psi_{\text{JIPDAS}}(h,g)$ & $\Psi_{\text{JIPDA}}(h,\dots,h,g)$ &  (\ref{SuperposedJIPDAS}) & $\le n$ pt. tgts. & PPP & Std. & Yes \Bstrut \\ \hline
		\bf{MHT}$^{2}$ & $\Psi_{\text{MHT}}(h_1,\dots,h_{n+m},g)$ & $\Psi_{C}^{\text{PPP}}(g)\,\prod_{i=1}^n\Big(1-\chi_i+\chi_i\,\Psi_{\text{BMD}(i)}(h_i,g)\Big)$ &  (\ref{pgflForJIPDADataDriven}) & $\le n+m$  & PPP & Std. & No \Tstrut \Bstrut \\
		\ & \ & $\times \prod_{j=1}^m \Big(1-\gamma_j+\gamma_j\,\Psi_{\text{BMD}(j)}^{\text{Data}}(h_{n+j},g)\Big)$ &  \ & pt. tgts. & \  & \ & \ \Bstrut \\	\hline
		\bf{Multi-Bernoulli} & $\Psi_{\text{MB}}(h,g)$ & $\Psi_{C}^{\text{PPP}}(g)\,\prod_{i=1}^n\Big(1-\chi_i+\chi_i\,\Psi_{\text{BMD}(i)}(h,g)\Big)$	& (\ref{pgflMultiBernoulli}) & $\le n+m$  & PPP & Std.  & Yes\Tstrut \Bstrut \\
		\ & \ & $\times \prod_{j=1}^m \Big(1-\gamma_j+\gamma_j\,\Psi_{\text{BMD}(j)}^{\text{Data}}(h,g)\Big)$  &  \ &  pt. tgts.  & \ & \ & \  \Bstrut \\ \hline
		\bf{PHD} & $\Psi_{\text{PHD}}(h,g)$ & $\Psi_{C}^{\text{PPP}}(g)\,G_N^{\text{PPP}}\big(\Psi_{\text{BMD}}(h,g)\big)$	 & (\ref{pgflPHDOriginal}) & PPP & PPP & Std. & Yes\Tstrut \Bstrut \\\hline
		\bf{CPHD} & $\Psi_{\text{CPHD}}(h,g)$ & $\Psi_{C}^{\text{Cluster}}(g)\,G_{N}^{\text{Cluster}}\big(\Psi_{\text{BMD}}(h,g)\big)$	& (\ref{PGFLCPHD}) & Cluster & Cluster & Std. & Yes\Tstrut \Bstrut \\\hline
		\bf{Generalized PHD} & $\Psi_{\text{GenPHD}}(h,g)$ & $\Psi_{C}^{\text{gen}}(g)\,G_N\big(\Psi_{\text{BMD}}^{\text{gen}}(h,g)\big)$	 & (\ref{pgflGeneralizedPHD}) & Gen.$^7$ & Gen.$^6$ & Gen.$^4$ & Yes\Tstrut \Bstrut \\ \hline	
		\bf{Joint PHD} & $\Psi_{\text{JointPHD}}(h_1,\dots,h_n,g)$ &  $\Psi_{C}^{\text{PPP}}(g)\,\prod_{i=1}^n G_{N_i}^{\text{PPP}}\big (\Psi_{\text{BMD}}(h_i,g)\big)$ & (\ref{jointPHDPGFL}) & $= n$ gps.   & PPP & Std.$^3$  & Yes$^3$ \Tstrut \Bstrut  \\  \hline 
		\bf{Joint Gen. PHD} & $\Psi_{\text{JointGenPHD}}(h_1,\dots,h_n,g)$ &  $\Psi_{C}^{\text{gen}}(g)\,\prod_{i=1}^n G_{N_i}^{\text{PPP}}\big (\Psi_{\text{BMD}}^{\text{gen}}(h_i,g)\big)$ &  (\ref{pgflJointGeneralizedPHD}) & $= n$ gps.  & Gen.$^6$ & Gen.$^{3,4}$  & Yes$^3$ \Tstrut  \Bstrut\\ \hline		
	\end{tabular}
	\begin{tabular}{llllllllll}
		\	&  \ & \ & \ &  \ & \ & \  & \ & \ & \ \\
	\end{tabular}
	\caption{Overview of the pointillist family of multitarget tracking filters}

	1 Clutter is modeled as target with high variance \ \ \ 2 MHT of un-pruned and complete set of hypothesis\ \ \ 3 Within target groups \ \ \ 4 Arbitrary target-oriented measurement process \ \ \ 5 Extended target measurement model
	 \ \ \ 6 General finite point process clutter model \ \ \ 7 General finite point process target model \ \ \ 8 Modeled via missing data assignment weight %(Expectation-Maximization algorithm)	
\end{table*}

\section*{Acknowledgements}
The first author thanks Prof. Juan Manuel Corchado Rodriguez of the University of Salamanca for inviting him to give the keynote lecture \cite{StreitFusion14Talk} that provided the impetus for this paper. He also thanks Profs. Taek Lyul Song and Darko Musicki (now deceased) of Hangyang University for their support and many stimulating technical discussions that led to the insights documented in \cite{MusickiSongStreit}. 

%% file: appendixA.tex
\label{sec:AppendixA}
The results given in this appendix, together with supporting technical details, can be found (in different notations and naming conventions) in \cite{DaleyJonesVolume1}, \cite{DaleyJones}, \cite{Moyal},  \cite{StreitPGFL}, \cite{MahlerBook}, \cite{NotesSummerSchool}, and the references cited therein. The latter two references, and the extensive literature cited in them, discuss a closely related version of the theory of finite point processes called finite set statistics (FISST) and random finite sets (RFSs). \\
\indent Let $\mathcal X$ be a separable metric space. A typical choice for $\mathcal X$ is $\mathbb R^d$, $d>0$, which is sufficient for the most applications appearing in target tracking. %(In this paper, $\mathcal X$ is either the target state space or the measurement space, depending on context.) 
The space of sets of points or event space in $\mathcal X$ is defined by
\begin{align}
	\label{EZ}
	\mathcal E_{\mathcal X}:= \emptyset\cup\bigcup_{n\ge1}\mathcal X{(n)},
\end{align}
where $\mathcal X{(n)}$ is the space of sets of size $n\in\mathbb N$, that is 
\begin{align}
	\mathcal X{(n)} := \left\{\{x_1,...,x_n\}\big{|}x_i\in \mathcal X, i=1,...,n\right\}.
\end{align} 
It is assumed that each element $\varphi\in \mathcal E_{\mathcal X}\setminus\emptyset$ is locally finite, i.e., each bounded subset of $\mathcal X$ can contain only a finite number of points of $\varphi$, and is simple, i.e.,
\begin{align}
	\forall x_i,x_j\in\varphi,\ \ x_i=x_j \Rightarrow i = j.
\end{align}
In physics (thermodynamics), $\mathcal E_{\mathcal X}$ is called the grand canonical ensemble, $\mathcal X{(n)}$ is called the $n^\textnormal{th}$ canonical ensemble and $n$ is called the canonical number. In terms of target tracking this translates to the assumptions that only finitely many targets can be present in a scenario and that no two targets share the same state. In tracking the number of targets is the canonical number and is usually referred to as the cardinal number. \\
\indent The definition of the grand canonical ensemble must be modified to accommodate applications in which the state space $\mathcal X$ is discrete or continuous-discrete. These kinds of spaces arise occasionally in tracking; for example, see \cite{StreitStoneFusion2008}, \cite{StoneStreitBMTT}, and \cite{StreitDiscreteIntensity}. In such spaces the discrete elements of $\mathcal X$ are allowed to be repeated, i.e., the sets in the grand canonical ensemble are allowed to be multisets. \\
\indent A stochastic process in the sense of \cite{Stoyan} is defined as a measurable mapping
\begin{align}
	\Phi:(\Omega,\mathcal F,\mathbb P)\to(\mathcal E_{\mathcal X},\mathcal B(\mathcal E_{\mathcal X})),
\end{align}
where $(\Omega,\mathcal F,\mathbb P)$ is an arbitrary probability space and $\mathcal B(\mathcal E_{\mathcal X})$ denotes the Borel $\sigma$--algebra of $\mathcal E_{\mathcal X}$. 
Let $P_\Phi$ denote the pushforward (image) measure of $\mathbb P$, using the point process $\Phi$. For any bounded and Lebesgue-integrable function
\begin{align}
	h:(\mathcal X,\mathcal B(\mathcal X)) \to (\mathbb R,\mathcal B(\mathbb R))
\end{align}
the PGFL $\Psi\equiv\Psi_\Phi$ of the point process $\Phi$ is defined by
\begin{align}
	&\Psi(h) \equiv \sum\limits_{n\ge 0} \int\limits_{\mathcal X(n)} \prod\limits_{i=1}^n h(x_i)P_\Phi(d\{x_1,\dots,x_n\})\\
	\label{Radon}
	&=\sum\limits_{n\ge 0} \frac 1 {n!} \int\limits_{\mathcal X^n}\prod\limits_{i=1}^n h(x_i)p_n(x_1,\dots,x_n)\,dx_1\cdots dx_n,
\end{align}
where $p_n:\mathcal X^{n}\to\mathbb R$ is the joint probability density of the corresponding Jannossy measure \cite{DaleyJonesVolume1} and defined such that
\begin{align}
	\int_B n!P_\Phi(d\{x_1,\dots,x_n\})=\int_B p_n(x_1,\dots,x_n)\,dx_1\cdots\, dx_n
\end{align} 
holds for all $B\in \mathcal B(\mathcal E_\mathcal X)$. Equation (\ref{Radon}) holds due to the assumed absolute continuity of $P_\Phi$ and the application of the Radon-Nikodym Theorem. Note that $P_\Phi(d\{x_1,\dots,x_n\})$ is the probability of the ordered event, and $p_n(x_1,\dots,x_n)\,dx_1\cdots dx_n$ is the probability of the unordered event.\\ %In \cite{Moyal} PGFLs were defined for general finite point processes as a generalization of probability generating functions (PGFs) for multivariate discrete random variables. There, it is also shown that PGFLs characterize the corresponding point process via its functional derivative.\\
\subsubsection{G\^{a}teaux derivative with respect to the Dirac delta}
\indent Let $\Psi$ be a PGFL defined as in (\ref{Radon}). Then the G\^{a}teaux derivative of $\Psi$ with respect to the variation $\omega$ is defined by
\begin{align}
	\label{derivative}
	\frac{\partial\Psi (h)}{\partial\omega} \equiv \underset{\epsilon\searrow0}{\lim} \frac{\Psi(h+\epsilon \omega)-\Psi(h)}{\epsilon},
\end{align}
where $\omega$ is a real-valued, bounded and Lebesgue-integrable function on $\mathcal X$. In \cite{Dirac1927} the Dirac delta is defined to be a function which satisfies $\delta(x) = 0 \text{ when }x\ne 0$ and $\int\delta(x)dx=1$. Due to the property of the Lebesgue integral that it does not depend on changing the integrand at one point, $\delta$ cannot be a proper function. See \cite{EinfDistr} for a proper handling of Dirac delta as a distribution. Alternative references are \cite{Strichartz} and \cite{Friedlander}. Dirac delta evaluated at the point $c$ is defined by $\delta^c(x) = \delta(x-c)$.\\
\indent The class of PGFLs of interest is defined by 
\begin{align}
	\label{PGFLsOfInterest}
	\mathcal P_2&\equiv \bigg\{\Psi:\mathcal H\to\mathbb R\,\Big{|} \nonumber\\
	&\Psi(h)=\sum\limits_{n\ge 0} \frac {a_n} {n!} \int\limits_{\mathcal X^n}\prod\limits_{i=1}^n h(x_i)\, p_n(x_1,\dots,x_n)\,dx_1\cdots dx_n\bigg\},
\end{align}
where 
\begin{align}
	\label{H}
	\mathcal H\equiv\{h:\mathcal X\to\mathbb R \big{|}h\text{ is bounded and Lebesgue-integrable}\},
\end{align}
and $p_n:\mathcal X^n\to\mathbb R$ is a symmetric PDF.\\
\indent For a family of test functions $\{\gamma_{\lambda}^c\}_{\lambda>0}$ of $\delta^c$, that is a family of functions being an approximate identity in the sense of \cite[p.114]{Alt}, the functional derivative with respect to an impulse $c\in \mathcal X$ is defined by 
\begin{align}
	\label{definitionGateauxDerivative}
	\frac{\partial \Psi}{\partial c}(h)\equiv\underset{\lambda\searrow 0}{\lim}\frac{\partial \Psi}{\partial \gamma_\lambda^c}(h),
\end{align}
where
\begin{align}
	\label{funcDerivative}
	\frac{\partial \Psi}{\partial h}[h']\equiv \underset{\epsilon\searrow 0}{\lim}\frac{\Psi(h+\epsilon h')-\Psi(h)}{\epsilon}
\end{align}
and $h,\,h'\in\mathcal H$. It can be shown by the application of the Lebesgue Dominated Convergence Theorem \cite{Alt} that the G\^{a}teaux derivative with respect to the Dirac delta (\ref{definitionGateauxDerivative}) is well-defined for the PGFLs from (\ref{PGFLsOfInterest}) \cite{Paper1}. \\
\indent The derivative with respect to multiple real-valued, bounded and integrable variations $\omega_1,\dots,\omega_m$ is defined iteratively, that is, by applying either (\ref{derivative}) or (\ref{definitionGateauxDerivative}) iteratively: 
\begin{align}
\delta^m \Psi(h;\omega_1,\dots,\omega_m) = \delta\left(\delta^{m-1}\Psi(h;\omega_1,\dots,\omega_{m-1});\omega_m\right).
\end{align}
Alternatively, the method of simultaneous perturbation can be chosen to define the functional derivative with respect to several variations \cite[Eq. (4.11)]{Moyal}. This method is used to define secular functions in Section \ref{sec:defineSecularFns}. 
\subsubsection{Factorial Moments and Summary Statistics}
\label{LinearMomentsandPDFS}
\indent The first factorial moment, or intensity function, of the point process is defined by the functional derivative of its PGFL $\Psi(h)$ with respect to the impulse $x\in\mathcal X$ by
\begin{align}
	\label{firstFactorialMoment}
	m_{[1]}(x) \equiv \frac{\partial \Psi}{\partial x}(1).
\end{align}
For PPPs it is straightforward to verify that the first factorial moment is the intensity function that parameterizes it. \\
\indent Higher order factorial moments are defined as higher order derivatives of $\Psi(h)$ evaluated at $h(x)=1$ \cite{DaleyJonesVolume1}. Thus, the $n^{\textnormal{th}}$ factorial moment is the functional derivative of $\Psi$ with respect to the impulses $x_1,\dots,x_n\in \mathcal{X}$, that is,
\begin{equation}
\label{nthFactorialMoment}
m_{[n]}(x_1,\dots,x_n) = \frac{\partial^n \Psi}{\partial x_1\cdots\,\partial x_n}(1).
\end{equation}
Higher order moments are used in tracking applications \cite{Bozdoganetal} to characterize the pair-correlation structure \cite{Stoyan} of the Bayes posterior process before closing the Bayes recursion. This important topic is outside the scope of the present paper. \\
\indent The first factorial moment, or intensity, for multivariate PGFLs is an extension of the definition (\ref{firstFactorialMoment}). Let the test functions $h_j:\mathcal X_j\to\mathbb R$,\ $j=1,\dots,n$ be non-negative and bounded, and let $x=(x_1,\dots,x_n)^T\in\mathcal X_1\times\cdots\times\mathcal X_n$. The first factorial moment of a multivariate PGFL $\Psi(h_1,\dots,h_n)$  is defined as the mixed first-order partial derivative 
\begin{align}\label{multivariateIntensity}
	m_{[1,\dots,1]}(x) \equiv \frac{\partial \Psi(h_1,\dots,h_n)}{\partial x_1\cdots\,\partial x_n}\Bigg \vert_{h_1\,=\,\cdots\,=\,h_n=1}.
\end{align}
The intensity is seen to be a multivariate function defined on $\mathcal X_1\times\cdots\times\mathcal X_n$. The higher order (mixed) factorial moments $m_{[k_1,\dots,k_n]}(x)$ can be defined analogously, as the mixed partial derivative of order $k_1,\dots,k_n$ with respect to the test functions $h_1,\dots,h_n$, respectively. \\
\indent The functional $\Psi(\cdot)$ is linear if, for all test functions $h,g$ and constants $a, b\in \mathbb{C}$,
\begin{equation}
\Psi(a h + b g) = a \Psi(h) + b \Psi(g).\label{linearFnal}
\end{equation}
It is straightforward to see that for linear functionals, the only \textit{nonzero} probability realizations of the point process have exactly one point, and the PDF of this point (in continuous spaces $\mathcal{X}$) is identical to the intensity (\ref{firstFactorialMoment}). Multivariate functionals are multilinear if they are linear in each test function separately. As in the univariate case, it is easily verified that, with probability one, realizations are singleton points (in a Cartesian product space), and the multivariate PDF is identical to the intensity function. \\  
%In \cite[Equation (5.4.12)]{DaleyJonesVolume1} the $n$th factorial moment is written intuitively as
%\begin{align}
%	\label{intuitiveFactorialMoment}
%	& m_{[n]}^\Phi(x_1,...,x_n) dx_1\cdots dx_n \nonumber\\
%	&= \text{Pr}\begin{pmatrix}
%	\text{ exactly one point of the process is}\\
%	\text{located in each infitesimal subset}\\ 
%	[x_i,x_i+dx_i), i=1,...,n 
%	\end{pmatrix},
%\end{align}
%which shows that the factorial moments can be interpreted as multi-point intensity functions if the points are distinct with probability one. \\%In case the point process is given by the superposition of $n$ independent point processes the $n$th factorial moment is given by 
%\begin{align}
%	m_{[n]}^\Phi(x_1,...,x_n) dx_1\cdots dx_n = m_{[1]}^\Phi(x_1)dx_1\cdots m_{[1]}^\Phi(x_n)dx_n.
%\end{align}\\
%The pair-correlation function is given by $m_{[2]}^\Phi(x_1,x_2)$. In tracking applications it characterizes the "spooky action" between two targets in the Bayes posterior process due to assignment interference. Multipoint-correlation functions can also be computed.\\
\indent The PGFL of the Bayes posterior process and its intensity function is now derived. Let $\mathcal X$ and $\mathcal Y$ be the target state and measurement space respectively. Typically $\mathcal X\subset\mathbb R^{d_1}$ and $\mathcal Y\subset\mathbb R^{d_2}$, $d_1, \ d_2>0$. Let $\Upsilon$ be a finite point process with events in $\mathcal E_{\mathcal Y}$, and let $\Xi$ be a finite point process with elements in $\mathcal E_{\mathcal X}$. Then analogously to (\ref{Radon}) the bivariate PGFL is defined on the product space $\mathcal E_{\mathcal Y}\times\mathcal E_{\mathcal X}$ as the expectation of the product of random products $\prod_{i=1}^m g(y_i)\prod_{j=1}^n h(x_j)$, that is, 
\begin{align}
	\label{bivariatePGFL}
	\Psi_{\Upsilon \Xi}(g,h) = \sum_{m=0}^\infty\sum_{n=0}^\infty\frac{1}{m!n!}\int_{{\mathcal Y}^m}\int_{\mathcal{X}^n}\prod_{i=1}^m g(y_i)\prod_{j=1}^n h(x_j) \nonumber\\
	\times\, p_{\Upsilon\Xi}(y_1,\dots,y_m,x_1,\dots,x_n)\,dy_1\cdots dy_m \,dx_1\cdots dx_n,
\end{align}
where $p_{\Upsilon\Xi}: \mathcal Y\times \mathcal X\to\mathbb R$ is a symmetric PDF in the first and second arguments, respectively.
Marginalizing with respect to one process yields the PGFL of the other, that is,
\begin{align}
	\label{MarginalizingBivariate}
	\Psi_{\Upsilon\Xi}(1,h) = \Psi_{\Xi}(h)\,\, \text{ and }\,\,\Psi_{\Upsilon\Xi}(g,1) = \Psi_\Upsilon(g).
\end{align}
Using functional derivative with respect to impulses and applying Bayes rule yields the PGFL of the Bayes posterior process (see details in \cite{StreitPGFL}), that is, 
\begin{align}
	\label{pgflOfBayesPosterior}
	\Psi_{\Xi|\Upsilon}(h|y_1,...,y_m) = \frac{\frac{\partial^m \Psi_{\Upsilon\Xi}}{\partial y_1\cdots\,\partial y_m}(0,h)}{\frac{\partial^m \Psi_{\Upsilon\Xi}}{\partial y_1\cdots \,\partial y_m}(0,1)}.
\end{align}
The ratio (\ref{pgflOfBayesPosterior}) for multi-target applications is given in \cite[p. 757]{MahlerBook}, where it is described as the PGFL form of the ``multi-target corrector.'' Conditional PGFs for discrete multivariate distributions take exactly the same form \cite[p. 11]{JohnsonKotz}.\\
\indent The intensity of the Bayes posterior process is often used as a summary statistic. For $x\in \mathcal X$, it is given by
\begin{align}
	\label{intensity}
	f^{\Xi|\Upsilon} (x)= \frac{\frac{\partial^m \Psi_{\Upsilon\Xi}}{\partial y_1\cdots\,\partial y_m\partial x}(0,1)}{\frac{\partial^m \Psi_{\Upsilon\Xi}}{\partial y_1\cdots \,\partial y_m}(0,1)}.
\end{align}
Another summary statistic is the posterior distribution of the canonical (cardinal) number, which is given by
\begin{align}
	\label{cardinalityDistribution}
	p_N^{\Xi|\Upsilon}(n) =\, &\frac 1 {n!} \frac{d^n}{dx^n} F^{\Xi|\Upsilon}(0)\nonumber\\
		=\, & \frac{\frac 1 {n!} \frac{d^n}{dx^n}\left(\frac{\partial^m \Psi_{\Upsilon\Xi}}{\partial y_1\cdots\,\partial y_m}(0,1)\right)_{x=0}}{\frac{\partial^m \Psi_{\Upsilon\Xi}}{\partial y_1\cdots \,\partial y_m}(0,1)},
\end{align}
where 
\begin{align}
	\label{PGFNumberOfTargets}
	F^{\Xi|\Upsilon}(z) = \frac{\frac{\partial^m \Psi_{\Upsilon\Xi}}{\partial y_1\cdots\,\partial y_m\,\partial x}(0,1)\Big|_{h(\cdot)=z}}{\frac{\partial^m \Psi_{\Upsilon\Xi}}{\partial y_1\cdots \,\partial y_m}(0,1)},
\end{align}
$z\in\mathbb C$, denotes the PGF of the number of targets. The expected number of targets is given by $\frac{d}{dx}F^{\Xi|\Upsilon}(1)$. \\ 
%\indent Since a tracking filter is completely characterized by its factorial moments a filter that can be derived via PGFLs is completely determined by the PGFL of the Bayes posterior process. Since this PGFL is given by a ratio of the functional derivatives of the joint PGFL $G^{\Upsilon\Xi}$ the intensity function and thus the essential behaviour of a filter, which is derivable via a PGFL, is determined by its joint PGFL $G^{\Upsilon\Xi}$. From Section \ref{sec:Bayes-MarkovFilter} to Section \ref{sec:multiBernoulliIntensityFilter} the PGFLs of many well-known filters are presented and their similarities/differences are discussed. The class of these filters is called the family of Pointillist filter.\\
%\indent Besides functional differentiation the definition of a secular function \cite{StreitSecular} offers the possibility to derive (\ref{intensity}) in different ways. Section \ref{sec:alternativesToFunctionalDifferentiation} considers several alternatives to functional differentiation.

%% file: appendixB.tex
\label{sec:stackedVsRFS}
The difference between an ordered stack of $n$ random vectors and a random set comprising the same $n$ vectors is explained in terms of superposition. The difference comes alive in tracking problems because target labels are arbitrary and the order of the vectors in the stacked vector is irrelevant. \\
\indent Denote the $n$ random variables by $X_i$ and their realizations by $x_i\in S$. The stacked random vector is the $n$-tuple $(x_1,\dots,x_n)$, while the random set is $\{x_1,\dots,x_n\}$. The fact that the order is irrelevant is often misinterpreted to mean that the stacked vector and the random set are somehow equivalent. Careful examination shows that the random set is properly conceptualized as the superposition of a joint point process whose realizations are those of the random vector variable $X\equiv (X_1,\dots,X_n)$ defined on the product space $S^n$. \\
\indent The difference is seen clearly by comparing PGFLs. The stacked random vector has the joint PGFL $\Psi(h_1,\dots,h_n)$ with $n$ test functions, each defined on $S$. Realizations of the joint point process comprise exactly one $n$-tuple (with probability one). Its joint PGFL is 
\begin{align}
	\Psi&(h_1,\dots,h_n)\nonumber\\
	&=\int_{S^n} \Big(\prod_{i=1}^n h_i(x_i)\Big) p_X(x_1,\dots,x_n) \,dx_1\dots dx_n,\label{stackedRandomVector}
\end{align}
where $p_X(x_1,\dots,x_n)$ is the joint PDF of the random vector $X$. In contrast the PGFL of the random set is the diagonal of this joint PGFL, namely
\begin{equation}
	\Psi_S(h)=\int_{S^n} \Big(\prod_{i=1}^n h(x_i)\Big) p_X(x_1,\dots,x_n) \,dx_1\cdots dx_n.\label{randomSet}
\end{equation}
It has only one test function, also defined on $S$. The PGFLs are different, so the processes are different as well. \\
\indent The PDF of the stacked vector is found by functional differentiation of (\ref{stackedRandomVector}) with respect to the test functions $h_i$, each with a single point mass at $x_i$. As expected, the derivative is $p_X(x_1,\dots,x_n)$. In contrast, the PDF of the random set is found by functional differentiation of (\ref{randomSet}) 
with respect to the test function $h$ with $n$ point masses, one at each $x_i$. It can be shown that the derivative is
\begin{equation}
	p(\{x_1,\dots,x_n\})=\sum_{\sigma\in \textnormal{Sym}(n)} p_X(x_{\sigma(1)},\dots,x_{\sigma(n)}),
\end{equation}
where $\textnormal{Sym}(n)$ is the set of permutations of $(1,\dots,n)$. This expression is a sum over all random vectors that yield the random set $\{x_1,\dots,x_n\}$. It can be derived by a tedious direct calculation, or by an easy exercise using the Cauchy residue theorem \cite[Appendix]{Fusion2015Streit}).  (For similar expressions related to measurement to target assignment problems, see \cite{Uhlmann}.)\\